\documentclass[%
reprint,
amsmath,amssymb,
aps,
floats,
superscriptaddress,
preprintnumbers
]{revtex4-2}

\usepackage{graphicx}
\usepackage{dcolumn}
\usepackage{bm}
\usepackage{hyperref}

\usepackage{color}
\usepackage{cancel}
\usepackage[normalem]{ulem}
\usepackage{AAS_macro}

\newcommand*\df{\mathop{}\!\mathrm{d}}
\newcommand{\n}{\nonumber}

\newcommand*\vrr{\mathbf{r}}
\newcommand*\Hm{\mathrm{H}_2}
\DeclareMathOperator{\sgn}{sgn}

\makeatletter
\newcommand*{\rom}[1]{\expandafter\@slowromancap\romannumeral #1@}
\makeatother

\usepackage{multirow}

\renewcommand{\arraystretch}{1.4}

\begin{document}
\preprint{TTK-22-29}

\title{Stochasticity of Cosmic Rays from Supernova Remnants and\\ the Ionization Rates in Molecular Clouds}

\author{Vo Hong Minh Phan}
\email{vhmphan@physik.rwth-aachen.de}
\affiliation{Institute for Theoretical Particle Physics and Cosmology (TTK), RWTH Aachen University, 52056 Aachen, Germany}

\author{Sarah Recchia}
\email{sarah.recchia@unito.it}
\affiliation{Dipartimento di Fisica, Universit\'a di Torino, via P. Giuria 1, 10125 Torino, Italy}
\affiliation{Istituto Nazionale di Fisica Nucleare, via P. Giuria, 1, 10125 Torino, Italy}

\author{Philipp Mertsch}
\email{pmertsch@physik.rwth-aachen.de}
\affiliation{Institute for Theoretical Particle Physics and Cosmology (TTK), RWTH Aachen University, 52056 Aachen, Germany}

\author{Stefano Gabici}
\email{gabici@apc.in2p3.fr}
\affiliation{Universit\'e de Paris, CNRS, Astroparticule et Cosmologie, F-75006 Paris, France}

\date{\today}

\begin{abstract}
Cosmic rays are the only agent able to penetrate into the interior of dense molecular clouds. Depositing (part of) their energy through ionisation, cosmic rays play an essential role in determining the physical and chemical evolution of star-forming regions. To a first approximation their effect can be quantified by the cosmic-ray induced ionization rate. Interestingly, theoretical estimates of the ionization rate assuming the cosmic-ray spectra observed in the local interstellar medium result in an ionization rate that is one to two orders of magnitude below the values inferred from observations. However, due to the discrete nature of sources, the local spectra of MeV cosmic rays are in general not representative for the spectra elsewhere in the Galaxy. Such stochasticity effects have the potential of reconciling modelled ionization rates with measured ones. Here, we model the distribution of low-energy cosmic-ray spectra expected from a statistical population of supernova remnants in the Milky Way. The corresponding distribution for the ionization rate is derived and confronted with data. We find that the stochastic uncertainty helps with explaining the surprisingly high ionization rates observed in many molecular clouds. 
\end{abstract}

\maketitle

\section{Introduction}
\label{sec:intro}

Cosmic rays are generally believed to play an important role in determining the dynamics of star-forming regions \citep{wurster2018,padovani2020,semenov2021} and, more importantly, the complex chemistry in molecular clouds (MCs) \citep{dalgarno2006,caselli2012}. This is because UV photons and X-rays are so effectively shielded by these clouds \citep{mckee1989,krolik1983,silk1983} leaving CRs as the only capable agent to penetrate and ionize their interior. In this respect, the impact of CRs on MCs can be quantified by the so-called CR induced ionization rate \cite{padovani2020,gabici2022}. Observations of many different chemical species via their emission or absorption lines in the infrared or millimeter domains indicate ionization rates roughly within the range between $10^{-17}$ s$^{-1}$ and $10^{-15}$ s$^{-1}$ \cite{caselli1998,indriolo2012,neufeld2017}.

Theoretical estimates of the ionization rate require essentially two ingredients: {\it i)} the spectra of low energy CRs in the interstellar medium (ISM) surrounding MCs, and {\it ii)} the transport model to describe the penetration of CRs into the clouds. In fact, the first attempts to calculate the ionization rate simply ignored the effect of CR propagation into clouds \citep{hayakawa1961,spitzer1968,nath1994,webber1998} and, thus, their results depend only on the low-energy ($E<1$ GeV) CR spectra extrapolated from data at higher energies. This has been done most notably in Ref. \citep{spitzer1968} where the authors obtained the ionization rate of about $10^{-17}$ s$^{-1}$ (commonly known as the {\it Spitzer} value). Puzzlingly, this is actually at the lower end of the range typically inferred in MCs. 

Subsequent studies provided a better description for the transport of CRs into MCs and considered also the role of energy losses due to CR interactions in dense and neutral media. A natural starting point is the scenario where CRs move ballistically along magnetic field lines into MCs \cite{padovani2009}. Such a scenario maximizes the penetration of CRs and, thus, yields the most optimistic estimate for the ionization rate. However, it has been suggested that the gradient of CR density from outside to inside clouds due to energy losses might excite self-generated magnetic turbulence in the vicinity of MCs and hinder the transport of CRs into MCs \citep{skilling1976,cesarsky1978,morfill1982,morlino2015,schlickeiser2016,phan2018,ivlev2018}.  

In fact, the exact choice for the model of CR transport into MCs might depend strongly on the properties of clouds and their surrounding media. Interestingly, both models fail to explain the ionization rate measured for many clouds in the Galaxy assuming the spectra of low-energy CRs around these clouds to be similar to the ones observed locally by the Voyager probes \citep{cummings2016,phan2018,padovani2020,gabici2022}. This problem might be resolved by arguing that the Voyager spectra are local and, thus, there might exist regions of the ISM with higher density of low-energy CRs to induce the ionization rate typically observed \cite{silsbee2019,phan2021}. Indeed, such an inhomogeneity for low-energy CRs is expected since the relatively short energy loss time of these particles would not allow them to travel far away from their sources (see one of the first attempts to quantify this effect in Ref.~\cite{cesarsky1975}). Thus, a better understanding of the CR induced ionization rate might actually require the modelling of the interstellar transport of CRs.

Observational evidence such as the data of secondary-to-primary ratio of CR intensities indicates that the transport of GeV CRs within the Galactic disk is diffusive. There exists also observational constraint which comes from the analyses of unstable secondaries such as the radioactive isotope $^{10}$Be whose life time is roughly of the same order as the residence time of CRs inside the Galactic disk $\tau(^{10}\text{Be})\simeq 1.4$ Myr. The decay of this isotope would suppress the ratio between $^{10}$Be and stable Be isotopes by a factor of approximately $\tau(^{10}\text{Be})/\tau_{esc}$. Measurements of the Beryllium ratio then allow us to infer $\tau_{esc}\simeq 10-20$ Myr (e.g. \cite{gabici2019,evoli2020}). Since this timescale is much larger than the residence time within the Galactic disk, it is believed that CRs are also diffusively confined within a low-density magnetized halo surrounding the Galactic disk. Although the size of this halo is quite poorly constrained, observations of the diffuse synchrotron emission above and below the Galactic plane in the radio domain seem to indicate the thickness of the halo to be much larger than that of the Galactic disk \citep{beuermann1985}.

In the framework of the Galactic halo model, the standard approach to estimate the intensities of CRs, especially in the hadronic sector at high energy, is to model all the sources together as a smooth disk with a continuous CR injection such that the problem could be treated as steady-state \citep{strong1998,recchia2016,evoli2019,mertsch2021,genolini2021}. However, as discussed above, the short energy loss time of low-energy CRs could result in a more patchy distribution for these particles such that the contribution from recent or nearby sources might become crucial and, thus, the use of a continuous disk of sources might be no longer valid. In other words, the values for the intensities of low-energy CRs at different positions and times in the Galaxy can only be predicted if the exact locations and ages for all the CR sources are known. This is indeed not possible even for the most well-studied class of Galactic sources like SNRs since young and distant or very old SNRs are quite hard to observe even though these sources might contribute significantly to the CR intensities. 

Nevertheless, if the temporal and spatial \emph{distributions} of sources are assumed to be known, e.g. from the extrapolation of current surveys, the CR intensities could be evaluated for different realizations of possible locations and ages of the sources. The variations of these intensities from one realization to another are commonly referred to as stochasticity or stochastic fluctuations of CRs \cite{mertsch2011,blasi2012_stochastic,bernard2012,genolini2017,mertsch2018,manconi2020,evoli2021a,evoli2021b,phan2021}. Consequently, all the predictions for intensities or ionization rates are only probabilistic. More importantly, the stochasticity of low-energy CRs would allow us to identify also the most probable range of values for the ionization rate inside MCs. 

Concerning CR sources, it is not yet clear which classes of sources are the most dominant for Galactic CRs, especially in the MeV energy range that is most relevant for the process of ionization. Potential sources include (but are not limited to) wind termination shocks of stars or star clusters \cite{casse1980,cesarsky1983,scherer2008,morlino2021}, superbubbles \cite{parizot2004,bykov2014,bykov2020,tatischeff2021,vieu2022b,vieu2022c}, protostellar surface and jet shocks \cite{padovani2015,padovani2016,gaches2018,araudo2021,cabedo2022}, regions of turbulent reconnection inside molecular clouds \cite{gaches2021}, and SNRs \cite{tatischeff2018,gabici2019}. 
Interestingly, there is evidence of enhanced ionization rates in a few systems of SNR-MC associations which are also observed in GeV and TeV gamma-rays indicating the presence of CRs in the MeV and GeV energy range originating from these SNRs \citep{indriolo2010,ceccarelli2011,vaupre2014,gabici2015,phan2020}. It is for this reason that we shall focus only on the stochastic fluctuations due to SNRs assuming that they are the main sources of Galactic CRs.

We note also that the effect of stochasticity, in a sense, represents fluctuations of CR density on small scales and this  introduces a range of values for the ionization rate. There exists also variations of the CR density on larger scales, e.g. within spiral arms where the source density is much higher than in interarm regions. This should also result in an overall boost for the value of the ionization rate. Such variations can explain both the scatter in the observed ionization rate and their surprisingly high values.  

The paper is structured as follows. We shall first discuss the setup of the propagation model on Galactic scales for low-energy CR protons and electrons in the energy range from 1 MeV to about 10 GeV in Section \ref{sec:point-source}. In Section \ref{sec:stochasticity}, the stochasticity of the CR intensities will be analyzed at two representative points in our Galaxy for the distribution of source ages and distances taking into account the spiral structure of the Milky Way. The two representative models for the transport of CRs into MCs (the ballistic and diffusive models) shall be reviewed in Section \ref{sec:cloud} and the corresponding stochastic fluctuations of the ionization rate will be investigated in Section \ref{sec:ionization}. We summarize our results and conclude in Section \ref{sec:conclusion}.


\section{Galactic Cosmic-Ray Transport Model}
\label{sec:point-source}
Let's assume the transport of CRs from each source to be diffusive and isotropic such that the differential number density $\psi(\vrr,E,t)$ for CRs of kinetic energy $E$ at position $\vrr$ at time $t$ could be written as
\begin{eqnarray}
\psi=\sum_{i=1}^{N_s}\mathcal{G}(\vrr,E;\vrr_i,t-t_i),
\label{eq:psi}
\end{eqnarray}
where $\mathcal{G}(\vrr,E;\vrr_i,t-t_i)$ is the point-source solution (or Green's function) which, for each source $i$ in the disk, depends only on the kinetic energy $E$, the relative position from the point of interest to the source $\vrr-\vrr_i$, and the time of propagation since the injection of CRs $t-t_i$. The point-source solution can then be obtained by solving the following transport equation
\begin{eqnarray}
\frac{\partial \mathcal{G}}{\partial t}+\frac{\partial}{\partial z}\left(u\mathcal{G}\right)
-D\nabla^2\mathcal{G}+\frac{\partial }{\partial E}\left(\dot{E}\mathcal{G}\right)=q(\vrr,E,t),\label{eq:transport}
\end{eqnarray}
where $u=u(z)$ describes the advection profile with only the component perpendicular to the Galactic disk (this could be either advection by Galactic winds or streaming with the Alfv\'en speed), $D=D(E)$ is the isotropic and homogeneous diffusion coefficient, $\dot{E}$ describes the energy loss rate for CRs both inside the Galactic disk and in the magnetized halo, and $q(\vrr,E,t)$ is the source term which can be written as
\begin{equation}
q(r, z, E, t) = \sum_{i=1}^{N_\text{s}} Q(E)\delta(\vrr - \vrr_{i}) \delta(t-t_{i})\, .
\end{equation}
Here, we have to take into account the contribution from $N_s$ bursting sources located at $\vrr_{i}$ and releasing CRs at $t_{i}$ ($i=\left[1,N_s\right]$). Each of these sources injects CRs with the spectrum denoted as $Q(E)$ into the ISM. We shall not consider the effect of stochastic re-acceleration for reasons clarified below and the effect of this processes might be examined in future works. 
In the following, we discuss the specific form of all the physical quantities presented in the above equation in more detail.

\begin{table*}[!t]
\centering
\caption{Constrained and fitted parameters for the stochastic model for both CR protons and electrons in the case for the diffusion coefficient scaling with Lorentz factor $\gamma$ as $D\sim \beta \gamma^{\delta}$. We adopt the parameters from Ref. \cite{phan2021}, where they have been chosen in such a way that the stochastic model provides a good fit for the Voyager and AMS data assuming the distribution of sources as presented in Section \ref{sec:source-dist}.}

	\label{tab:parameters}
	\begin{tabular}{|c|c|c|r|} 
		\hline
		\hline
		\multirow{3}{*}{\shortstack{Fitted parameters\\ for low-energy CRs}} & $n_{\mathrm{H}}$ & Gas density in the disk & 0.9 cm$^{-3}$\\
		\cline{2-4}
		& $u_0$ & Advection speed & 16 km/s\\
		\cline{2-4}
		& $2h_s$ & Height of the disk of sources & 80 pc\\
		\hline
		\multirow{9}{*}{\shortstack{Constrained parameters\\ from high-energy CRs}} & $R_d$ & $\qquad$ Radius of the Galactic disk $\qquad$ & 15 kpc \\
		\cline{2-4}
		& $H$ & Height of the CR halo & 4 kpc\\
		\cline{2-4}
		& $2h$ & Height of the gas disk for energy loss & 300 pc\\
		\cline{2-4}
		& $D(E=10\textrm{ GeV})$ & Diffusion coefficient at 10 GeV & $5\times 10^{28}$ cm$^2$/s\\
		\cline{2-4}
		& $\delta$ & Index of the diffusion coefficient & 0.63\\
		\cline{2-4}
		& $\mathcal{R}_s$ & Source rate & 0.03 yr$^{-1}$\\
		\cline{2-4}
		& $\xi_{\text{CR}}^{(p)}$ & Proton acceleration efficiency & 8.7\%\\
		\cline{2-4}
		& $\xi_{\text{CR}}^{(e)}$ & Electron acceleration efficiency & 0.55\%\\
		\cline{2-4}
		& $\alpha$ & Index of the injection spectra & 4.23\\   
		\hline
		\hline
	\end{tabular}
\end{table*}

The advection profile is commonly modelled as $u(z)=u_0\sgn(z)$ \cite{jaupart2018}. Such an advection velocity might be due to the presence of a large-scale Galactic wind or, in the case where the magnetic turbulence is self-generated via the streaming instability, CRs would stream away from the Galactic disk at the Alfv\'en speed \citep{recchia2016}. In fact, one might expect higher-speed Galactic winds at large distance from the disk but more sophisticated advection profiles is not necessary. This is because advection is most important in modeling spectra of MeV to sub-GeV CRs. Since these low-energy particles probably never escape the disk due to energy loss, they are affected mostly by the advection speed around the disk $|z|\lesssim h$. In this region, it would be reasonable to expect advection to be due to Alfv\'en waves and, thus, $u_0$ should be of order $\sim 10$ km/s. We note also that our assumption of having advection due to Alfv\'en waves is motivated by models where Alfv\'en waves are self-generated by CRs \cite{blasi2012}. This is, in fact, compatible with neglecting the effect of stochastic re-acceleration in the transport equation (Eq. \ref{eq:transport}) as this process would require having Alfv\'en waves moving both upward and downward perpendicular to the disk which results in a small or vanishing value of the mean Alfv\'en speed \cite{evoli2019}.

Since the advection velocity changes its direction only at the Galactic disk, the term for the adiabatic energy loss could be taken into account as $\dot{E}=\dot{E}_{ad}=pvu_0/(3h)$ where $h$ is the half-thickness of the Galactic disk. Note also that, in this approximation, the adiabatic energy loss term acts everywhere within the disk ($|z|<h$). Such an approximation is quite standard in numerical treatments of Galactic CRs (see e.g. \cite{jaupart2018,evoli2019}).

Regarding the diffusion coefficient, it is not clear what should be the energy dependence of $D(E)$ at low energies. However, it was suggested from theoretical and observational analyses for particle transport in the solar system \citep{bieber1994,schlickeiser2010} that $D(E)\sim v_p\sim E^{0.5}$ in the non-relativistic limit since the particle mean free path becomes constant in that energy domain. Also, at high energy, the diffusion coefficient is expected to behave as $D(E)\sim E^{\delta}$ where $\delta\simeq 0.3-0.6$ (see \cite{strong2007,trotta2011} for more discussions). Combining the two asymptotic behaviors at high and low energy, we shall assume the following form for the diffusion coefficient:
\begin{eqnarray}
D(E)=D_0\beta\gamma^\delta,\label{eq:diff_ISM}
\end{eqnarray}
where $\gamma$ is the Lorentz factor and, for both CR protons and electrons, the index $\delta$ has been chosen to be $\delta\simeq 0.63$ similar to the results for high energy CRs from \cite{evoli2019}. We also normalize the diffusion coefficient such that $D(E=10\,{\rm GeV})\simeq 5\times 10^{28}$ cm$^2$/s for both CR protons and electrons. We note also that changing the form of the diffusion coefficient below $E=1$ GeV might not qualitatively affect the results at low energies as the intensity in this energy range is shaped by the process of energy loss, advection, and the source distribution in the vicinity of the observer (see e.g. the Supplemental Material of Ref.~\cite{phan2021} for a discussion concerning alternative forms of the diffusion coefficient).

The next element to be considered for the transport of CRs is the energy loss rate. Cosmic-ray protons lose energy mostly inside the Galactic disk via ionization and proton-proton interactions. Cosmic-ray electrons also suffer from energy loss inside the disk due to ionization and bremsstrahlung radiation. High-energy electrons, however, lose energy more efficiently via synchrotron radiation and inverse Compton scattering which occur not only in the disk but also in the magnetized halo. In practice, all of the above-mentioned mechanisms apart from synchrotron radiation and inverse Compton scattering are assumed to be effective only within the Galactic disk. We shall assume the vertical extent of the disk and the average hydrogen atom number density to be respectively $2h=300$ pc and $n_{\rm H}=0.9$ cm$^{-2}$ which corresponds to the disk surface mass density of about $2$ mg/cm$^3$ as inferred from observations \cite{ferriere2001}. We refer interested readers to the Supplemental Material of Ref.\,\cite{phan2021} and references therein for a more in-depth discussions on different mechanisms for energy loss. 

As mentioned above, we shall assume the dominant sources for Galactic CRs to be SNRs and adopt the injection spectrum as a power-law in momentum down to a kinetic energy of $1$ MeV
\begin{eqnarray}
Q(E)=\frac{\xi_{\text{CR}}E_{\text{SNR}}}{(mc^2)^2\Lambda\beta}\left(\frac{p}{mc}\right)^{2-\alpha},\label{eq:source_function}
\end{eqnarray} 
where $\xi_{\text{CR}}$ is the CR acceleration efficiency of the source, $E_{\text{SNR}}\simeq 10^{51}$ erg is the total kinetic energy of the supernova explosion, and $\Lambda$ is the normalisation given by the following integral
\begin{eqnarray}
\Lambda=\int^{p_{\text{max}}}_{p_{\text{min}}}\left(\frac{p}{mc}\right)^{2-\alpha}\left[\sqrt{\left(\frac{p}{mc}\right)^2+1}-1\right]\frac{\df p}{mc}.\label{eq:Lambda_Q}
\end{eqnarray}
The index of the source spectra is taken to be $\alpha=4.23$ which is compatible with the fit of observational data for high energy CR protons \cite{evoli2019}. In fact, this index is also achievable from nonlinear diffusive shock acceleration mechanism \citep{caprioli2012}. We also choose the efficiency of acceleration for CR protons and electrons as $\xi^p_{CR}\simeq 8.7\%$ and $\xi^e_{CR}\simeq 0.55\%$ respectively in order to match the observed data at high energy. 

Having established all the relevant ingredients, Eq. \ref{eq:transport} was solved numerically in the energy range from $1$ MeV to $10$ GeV within the region of transport defined above using an implicit finite difference scheme. Note that we have adopted the free-escape spatial boundary conditions $\mathcal{G}(r,z=\pm H,E,t)=0$ where $H$ represents the height of the CR halo and the radius of the Galactic disk respectively. 
All the parameters for the transport model of Galactic CRs are summarized in Table \ref{tab:parameters} which has been adopted from the Supplemental Material of Ref.\,\cite{phan2021}.

\section{Cosmic Rays from\\ Stochastic Sources}
\label{sec:stochasticity}

\subsection{Source Distribution}
\label{sec:source-dist}
As briefly mentioned above, we have to generate many different realizations of sources in order to estimate the stochastic uncertainty of the CR intensities. In this section, we shall discuss in more details the spatial and temporal distribution of SNRs. As a remark, we note that recent developments in the study of Galactic spiral structures, for example via data of molecular masers associated with very young high-mass star \cite{reid2019}, could provide a more updated view of the spatial source distribution, especially in the Galactic neighborhood. However, this is beyond the scope of the current work and might be investigated in the future.
\begin{figure*}[!t]
\centering
\includegraphics[width=1.03\columnwidth]{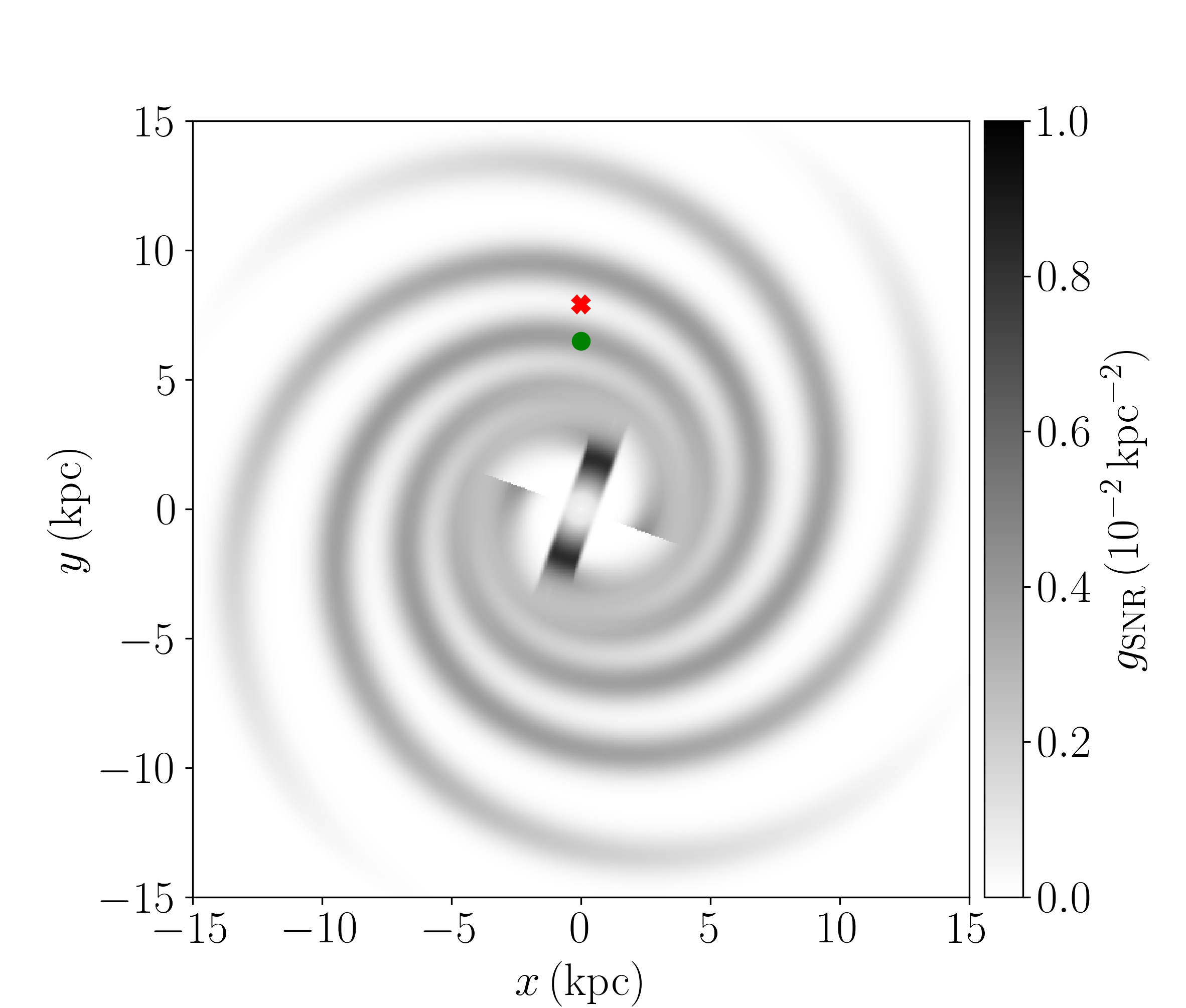}\hfill
\includegraphics[width=1.03\columnwidth]{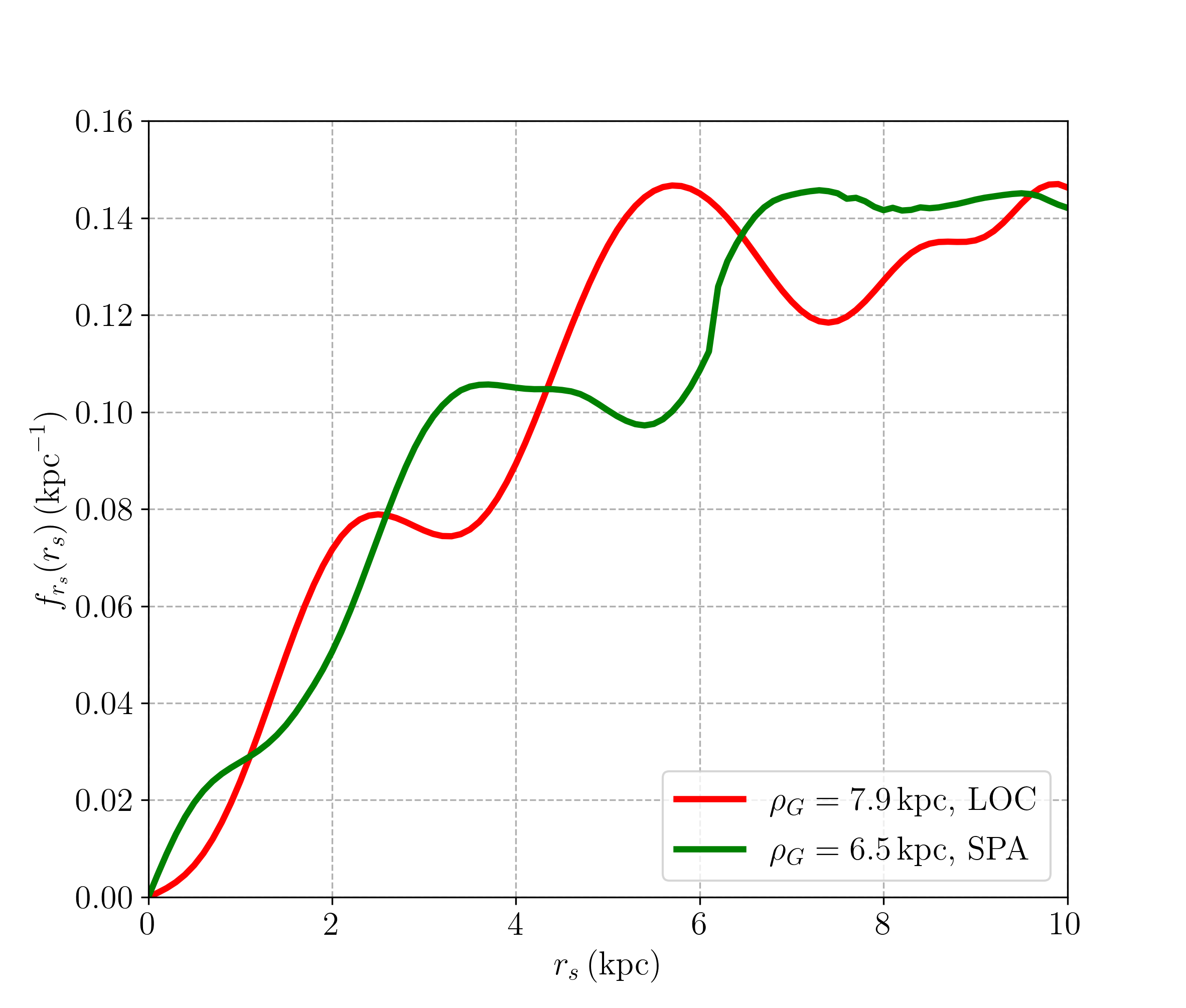}
\caption{{\bf Left:} The surface density of observed SNRs $g(\rho_G,\phi_G)$ in the Galaxy. The cross and filled circle marks respectively the position of the solar system and the observer at $\rho_G=6.5$ kpc in the spiral arms (referred to as the LOC and SPA point respectively, see text for more details). {\bf Right:} The probability density with respect to $r_s$ which is the distance between the observer and an SNR projected onto the midplane of the Galactic disk.}
\label{fig:SNR_density&spiral_structure}
\end{figure*}

In this work, we shall use the spatial distribution of SNRs $g (\rho_G, \phi_G)$ ($\rho_G$ and $\phi_G$ are respectively the Galactocentric radius and the Galactic azimuth) as presented in Ref.\,\cite{ahlers2009} (see also \cite{mertsch2011}). The authors have taken into account also the spiral structure of the Galaxy by adopting a logarithmic spiral with four arms of pitch angle $12.6^\circ$ and a central bar of 6 kpc length inclined by $30^\circ$ with respect to the direction Sun - Galactic centre \cite{vallee2005}. Besides, the density of SNRs for each arm is modeled by a Gaussian with 500 pc dispersion \cite{pohl1998}. 
We note also that $g (\rho_G, \phi_G)$ has been normalized with respect to $\phi_G$ in such a way that the surface density of observed SNRs in the Galaxy as a function of $\rho_G$ follows \citep{case1998}
\begin{equation}
f (\rho_G) = A \sin{\left(\frac{\pi \rho_G}{\rho_0} + \theta_0\right)} \text{e}^{-\beta \rho_G},
\label{eqn:SNR_density}
\end{equation}
where $A = 1.96\,\text{kpc}^{-2}$, $\rho_0 = 17.2\,\text{kpc}$, $\theta_0
= 0.08$ and $\beta = 0.13\,\text{kpc}^{-1}$. In the left panel of Fig.~\ref{fig:SNR_density&spiral_structure} we show the expected surface density of SNRs  $g (\rho_G, \phi_G)$.

Since we would like to analyze the stochastic fluctuations at different places in our Galaxy, it might be more straightforward to switch to the probability density in the coordinates centered on the position of interest denoted as $(r_s, \phi_s)$ where $r_s$ is the distance between the position of interest and an SNR projected onto the Galactic plane and $\phi_s$ is the polar angle. We can average over $\phi_s$ for simplicity (since the CR density from individual sources only depends on the absolute distance to the observer) and normalize the distribution to obtain the probability density $f_{r_s}$ with respect to $r_s$ as follows
\begin{equation}
 f_{r_s} (r_s) = \frac{\displaystyle\int_0^{2\pi} \text{d}\phi_s \, r_s \, g (\rho_G (r_s, \phi_s), 
 \phi_G (r_s, \phi_s))}{\displaystyle\int_0^{r_{\rm max}} \text{d}r_s\int_0^{2\pi} \text{d}\phi_s \, r_s \, g (\rho_G (r_s, \phi_s), 
 \phi_G (r_s, \phi_s))}.  
\label{eqn:dF}
\end{equation}
This function is shown in the right panel of Fig.~\ref{fig:SNR_density&spiral_structure} for two different cases where i) the observer is at the position of the solar system ($\rho_G\simeq 8$ kpc) and ii) the observer is within a spiral arm at $\rho_G=6.5$ kpc. We shall, from now on, refer to these two points respectively as the LOC (local) and SPA (spiral arm) positions and restrict ourselves to the analysis of the stochastic fluctuations at these two points. Here, we limit ourselves to $r_s\leq r_{\rm max}=10$ kpc since further sources do not contribute significantly if the standard explosion energy is assumed. 

It has been shown in some of the previous works that the vertical extension of sources might also affect the predicted intensities of CRs at low energies (see e.g. \cite{schlickeiser2014} for the smooth disk framework). Thus, we also consider a homogeneous vertical distribution of SNRs about the midplane of the Galaxy for the distribution of sources in the $z$-direction 
\begin{equation}
 f_{z_s} (z_s) = \left\{\begin{array}{ll} \dfrac{1}{2h_s\vphantom{\dfrac{}{}}} & \text{for } 
 |z_s| \leq h_s, \\ 0 & \text{otherwise}, \end{array}\right.
\end{equation}
We note that the vertical extensions might be different for different populations of supernovae. The vertical extensions of core-collapse and type Ia supernovae are expected to be respectively $2h_s\simeq 80$ pc and $2h_s\simeq 600$ pc \cite{prantzos2011}. Given the uncertainties on the vertical extension of source distribution for different populations of supernovae and the potential variations of these parameters on Galactic scale, we regard the half thickness of source distribution $h_s$ as an effective fit parameter representative for both populations core-collapse and type Ia supernovae. As we shall discuss later, this parameter is essential at low energies as it would partially determine the characteristic energy $E^*$ below which the effect of stochasticity is most important (see Eq. \ref{eq:Estar}). More important, this parameter also determines the average distance between SNRs and, thus, increasing $2h_s$ might lead to a smaller stochastic fluctuations. We note also that the rate of core-collapse supernovae is expected to be a few times higher than that of type Ia supernovae. This means that the value of $2h_s$ should be close to the expected value for core-collapse supernovae and, thus, we adopt $2h_s\simeq 80 \, \text{pc}$ for the vertical extension of sources in the following. 

The oldest sources that we consider are the ones that have released CRs around 100 millions years ago. This means that the time of propagation since the injection of CRs for all sources is $\tau=t-t_s\leq\tau_{max}=10^8$ yr which is longer than both the diffusive escape time of high-energy CRs and the energy loss time of low-energy CRs. Therefore, the obtained stochastic fluctuations are expected to reach a stationary state. We will further assume that the sources are uniformly distributed in time such that their probability density $f_{t_s}(t_s)$ is
\begin{equation}
 f_{t_s} (t_s) = \left\{\begin{array}{ll} 1/\tau_\text{max} & \text{for } 
 0 \leq t_s \leq \tau_\text{max}, \\ 0 & \text{otherwise}, \end{array}\right.
\end{equation}

Note that the stochastic fluctuations of the local low-energy CR spectrum, using the source distribution around the LOC point, has been recently investigated in Ref.\,\cite{phan2021}. In this work, we will extend the previous analyses and consider also the stochasticity at the SPA point. The corresponding stochastic fluctuations of the ionization rate shall be derived for both cases.

We could now build up a statistical ensemble by generating a large number $N_\text{r}=2000$ of realisations, in each drawing a large number of sources $N_\text{s}$ from the spatial and temporal distributions discussed above. In the previous analysis of the stochasticity for low-energy CRs around the solar system, we have estimated roughly the total number of discrete sources within $r_{\rm max}=10$ kpc in each realisation as $N_{s,\odot}=\mathcal{R}_s\tau_{\rm max}r^2_{\rm max}/R_d^2\simeq 1.33\times 10^6$ where the radius of the Galactic disk has been assumed to be $R_d\simeq 15$ kpc. If we place ourselves within the spiral arm at $\rho_G=6.5$ kpc, the number of sources within $r_{\rm max}$ is expected to be about 15\% higher as could be estimated from Eq.\,\ref{eqn:SNR_density} and $g(\rho_G,\phi_G)$. The total number of sources within $10$ kpc for an observer at $\rho_G=6.5$ kpc is then $N_s\simeq 1.52\times 10^6$. 

\subsection{Stochasticity of the Cosmic-Ray Spectra}
The CR intensities at an arbitrary position for each realization are evaluated by summing the contribution from all SNRs. In the following, we shall use the index $n$ to refer to a particular realization and each source within each realization will be associated with an index $i$. The intensity of CRs for a particular realization could be written following Eq.\,\ref{eq:psi} as
\begin{eqnarray}
j^{(n)}_{p,e}(E,t)=\frac{v}{4\pi}\sum_{i=1}^{N_s}\mathcal{G}(\vrr,E;\vrr^{(n)}_{i},t-t^{(n)}_{i}).
\end{eqnarray}
The index $p,e$ denote protons, electrons respectively, $v$ denotes the velocity of a particle at energy $E$. It follows that the expectation value for the intensities could be estimated as
\begin{eqnarray}
\langle j_{p,e}(E,t)\rangle=\frac{1}{N}\sum^{N}_{n=1}j^{(n)}_{p,e}(E,t).
\end{eqnarray}
We can now derive the probability density function for the value of the intensity and interpret the range of its most probable values. Note that these distribution functions are not symmetric and, more importantly, they do not have a well-defined second moment as has been shown for several analyses of the same type at high energy (see e.g. \cite{mertsch2011,bernard2012,genolini2017}). We shall, therefore, define the uncertainty intervals of the intensity using the percentiles (similar to that of \cite{mertsch2011}). Let's denote $p(j)$ as the probability density function of the CR intensity and define the uncertainty ranges using the percentiles \cite{mertsch2011}, e.g. $j_{a\%}$ could be determined via
\begin{eqnarray}
\int^{j_{a\%}}_{0}p(j)\df j=a\%.
\end{eqnarray} 
We could then identify the $95\%$ uncertainty ranges as $\mathcal{I}_{95\%}=\left[ j_{2.5\%},j_{97.5\%}\right]$. 

\begin{figure*}[htpb]
\includegraphics[width=3.5in, height=2.8in]{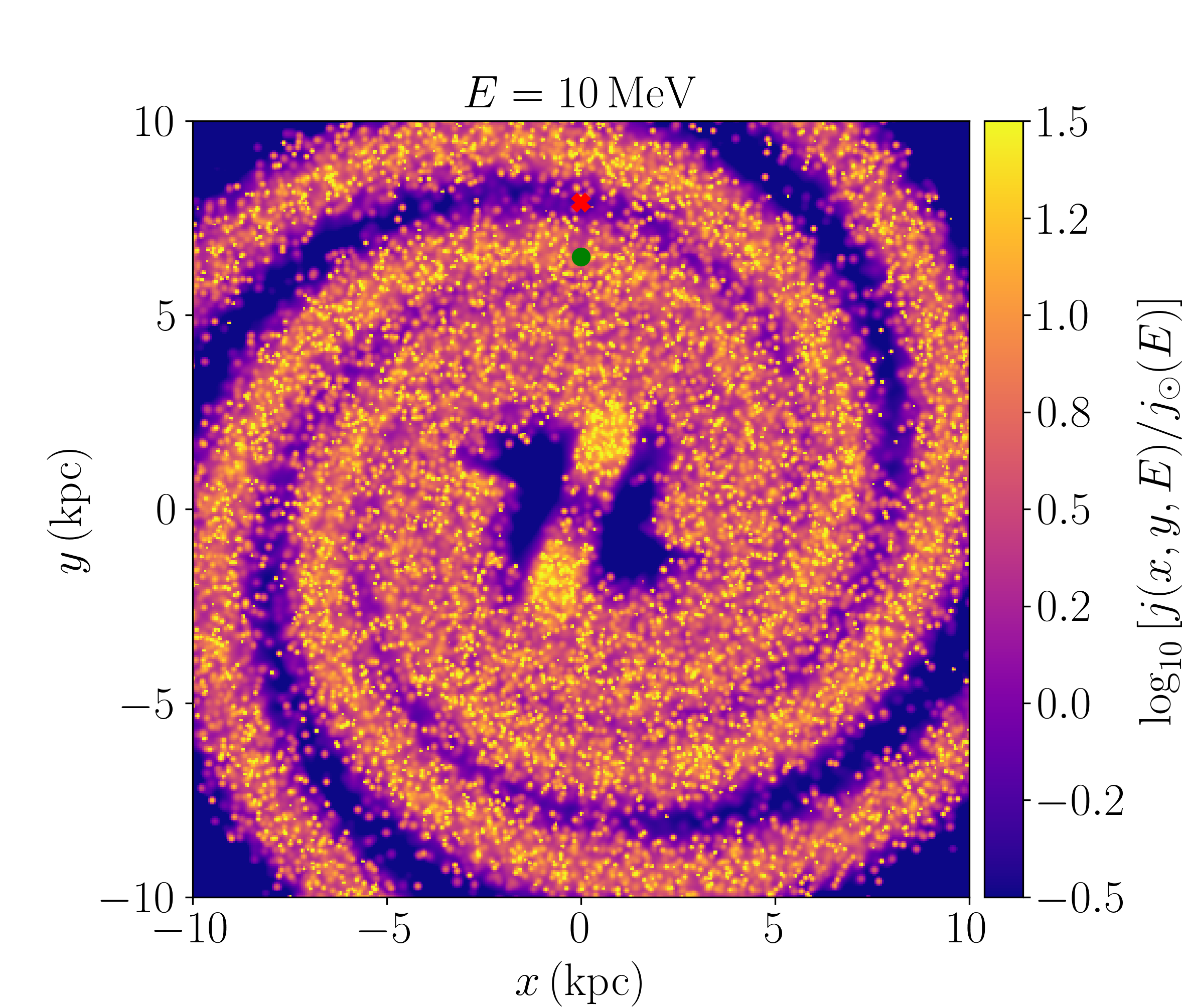}
\includegraphics[width=3.5in, height=2.8in]{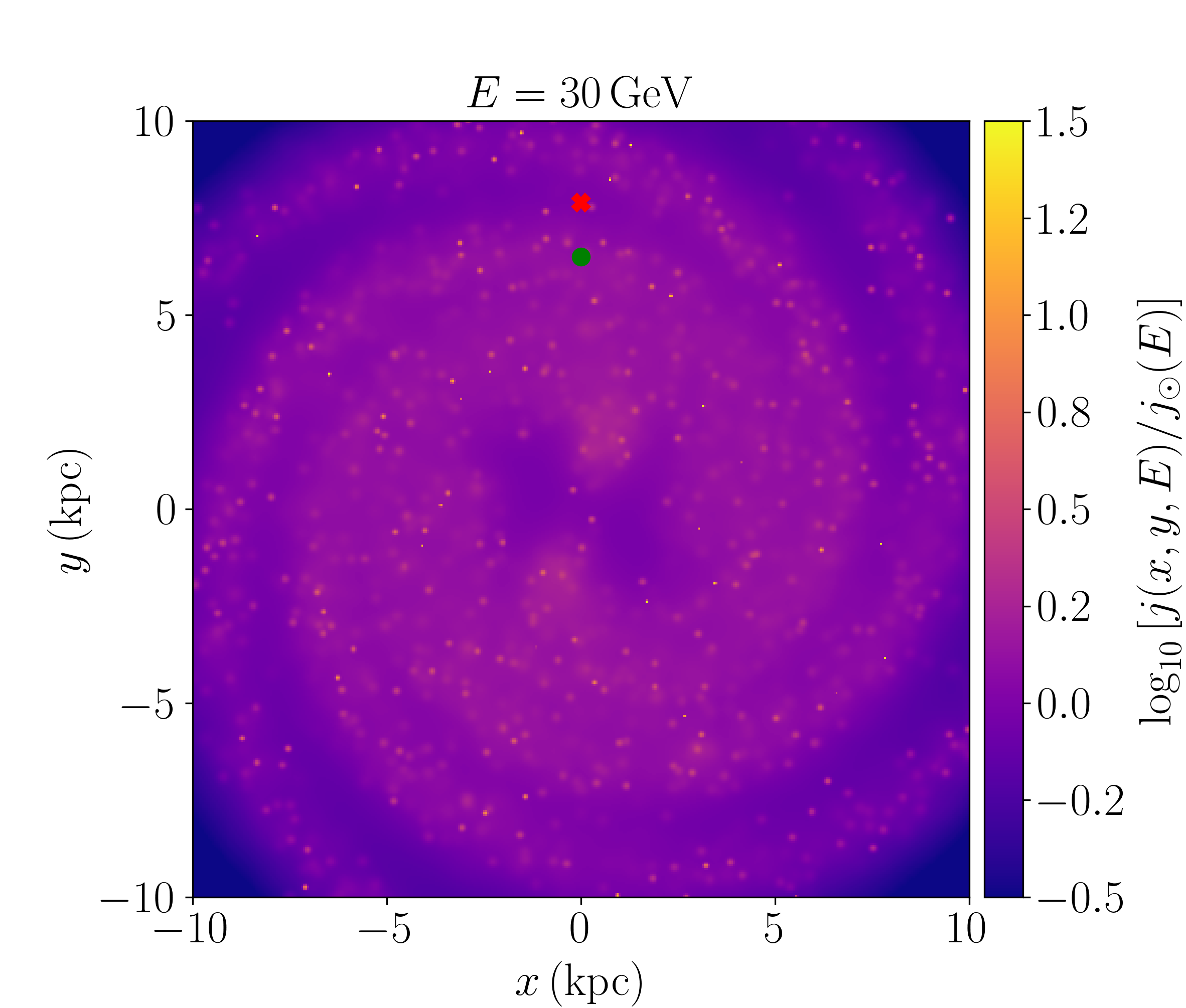}
\caption{Relative intensities of CR protons ($j/j_{\odot}$) at $E=10$ MeV (left panel) and $E=30$ GeV (right panel) in a particular realization of SNRs.}
\label{fg:jE/jsol}
\end{figure*}

Before looking into the stochastic fluctuations of the CR intensities at specific locations in the Galaxy, we would like to illustrate how CRs at different energies would distribute within a particular realization of SNRs. We present in Fig. \ref{fg:jE/jsol} the intensity of CR protons at $E=10$ MeV (left panel) and $E=30$ GeV (right panel) at different locations in the plane of the solar system $z=z_\odot\simeq14$ pc (divided by the intensity at the solar system, marked with the cross, in this realization). It is clear that the intensity of MeV CRs vary more significantly that of GeV CRs. In fact, the intensity of these low-energy particles is more patchy, i.e. peaking strong around their sources due to energy loss. This will ultimately lead to large stochastic fluctuations from one realization of SNRs to another as the intensities of MeV CRs are very sensitive to the exact locations of SNRs.

\begin{figure*}[htpb]
\includegraphics[width=3.5in, height=2.8in]{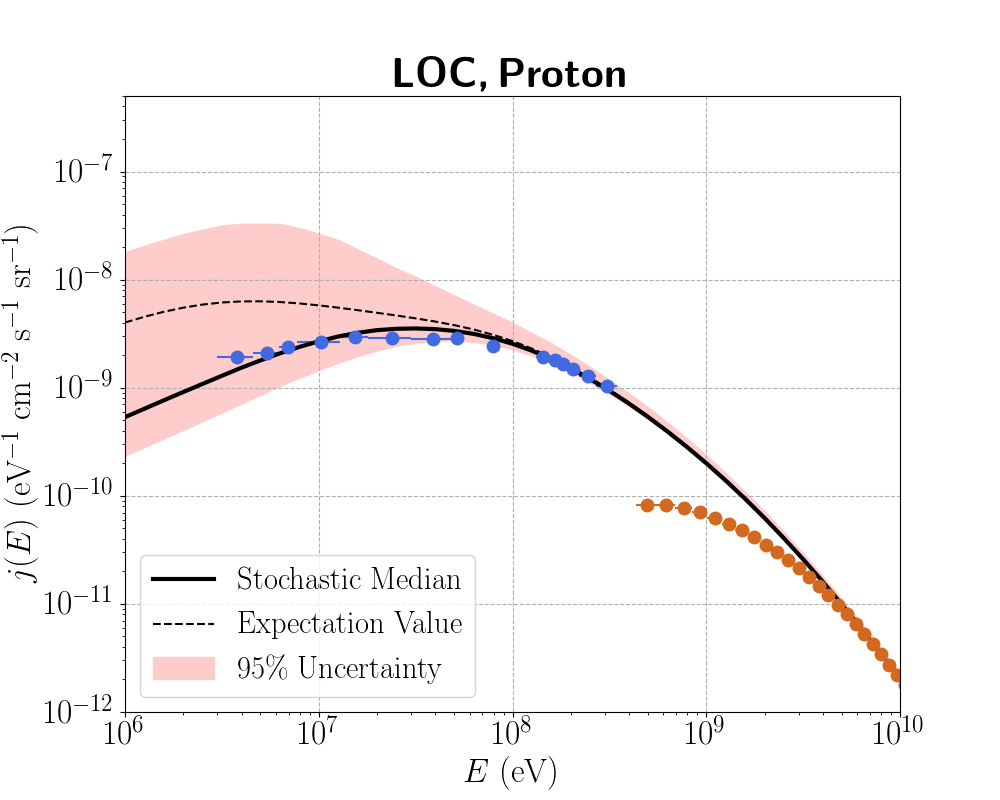}
\includegraphics[width=3.5in, height=2.8in]{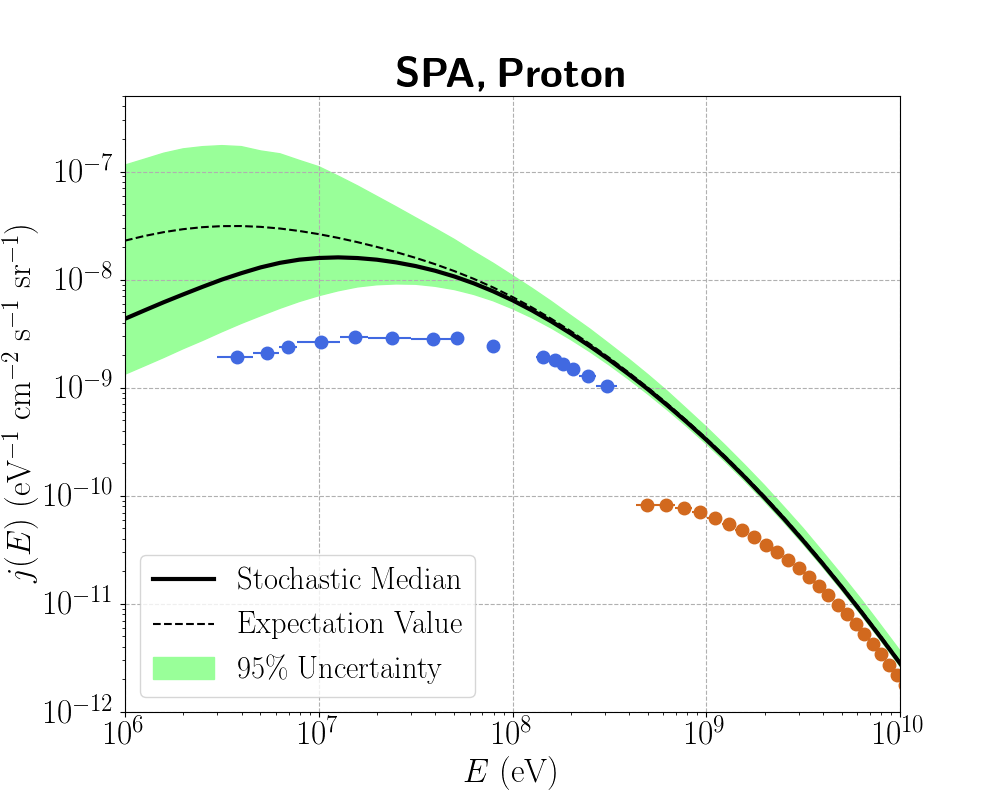}\\
\includegraphics[width=3.5in, height=2.8in]{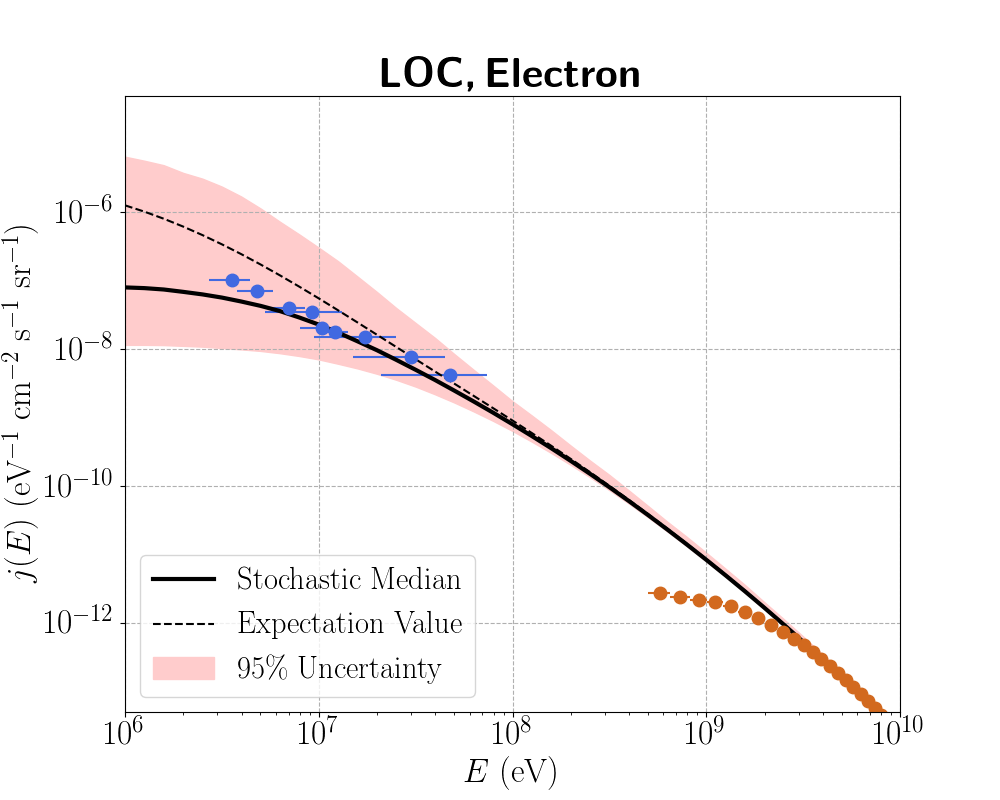}
\includegraphics[width=3.5in, height=2.8in]{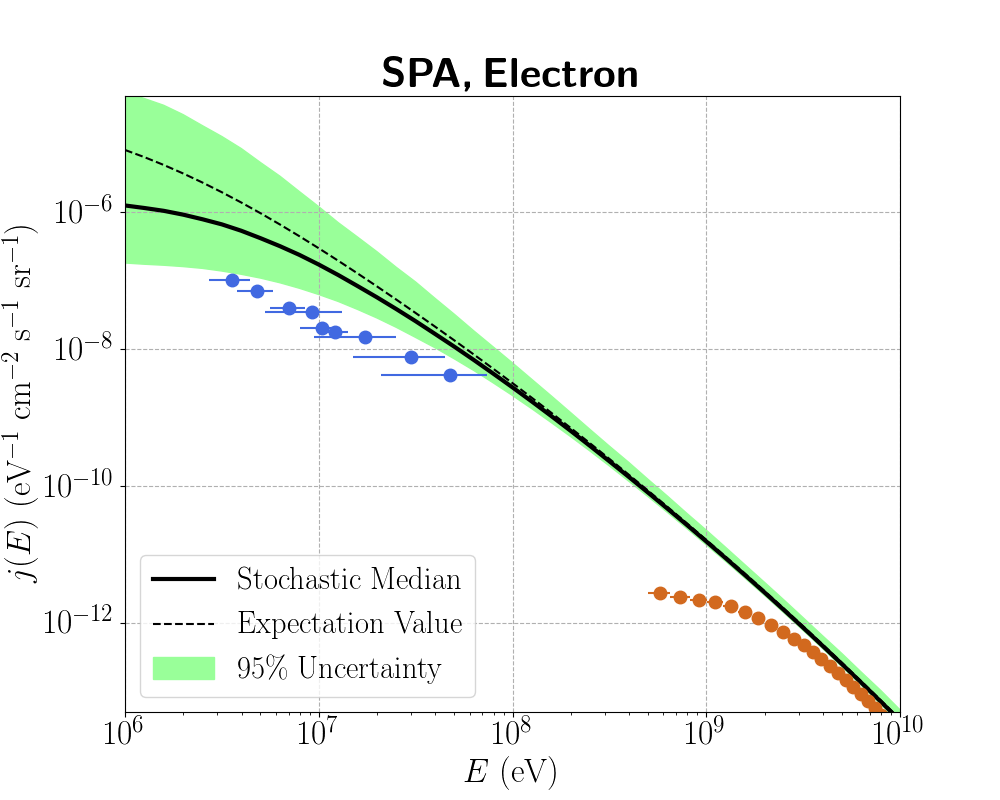}
\caption{Intensities and corresponding stochastic fluctuations of CR protons (upper panels) and electrons (lower panels) in comparison with data from Voyager~1~\cite{cummings2016} (blue) and AMS-02~\cite{AMS2014,AMS2015} (orange). Results are presented for an observer in the local ISM (the LOC position, left column) and in a spiral arm (the SPA position, right column). The dotted and solid black curves are respectively the expectation values and the median of the intensities. The shaded red or green regions are the 95\% uncertainty ranges of the intensities.
}
\label{fg:stochastic}
\end{figure*}

In Fig.~\ref{fg:stochastic}, the uncertainty ranges for the intensities of CR protons and electrons with energy from 1 MeV to about 10 GeV are presented as shaded regions together with the expectation values of the intensities and data from Voyager 1 \citep{cummings2016} and AMS-02 \citep{AMS2014,AMS2015}. 
Results are shown for two different positions in the Milky Way which are the LOC and SPA points as indicated in Section \ref{sec:source-dist} (see also the left panel of Fig.~\ref{fig:SNR_density&spiral_structure}). We have adopted the parameters for the sources and the transport of CRs that fit the locally observed intensities as in Ref.~\cite{phan2021} (see also Table \ref{tab:parameters}). Note also that the fit has been performed with the median intensities and not the expectation value of the intensities. This is because $\langle j(E)\rangle$ does not actually have the spectral behaviour expected for individual realizations.
In fact, the slope of the intensity below a characteristic $E^*$ follows $j(E)\sim v/\dot{E}$, as it has been pointed out in Ref. \cite{phan2021}. The authors also suggest a quite straightforward way to determine $E^*$. Roughly speaking, we expect a uniform distribution of CRs if the number of sources within the diffusion loss length $l_d(E)\simeq\sqrt{4D(E)\tau_l(E)}$ in the disk is much larger than one,
\begin{eqnarray}
\mathcal{R}_s\tau_l(E) \frac{2 l_d^3(E)}{3 R_d^2 h_s}\gg 1 \, .
\label{eq:Estar}
\end{eqnarray}
The characteristic energy $E^*$ can be estimated by setting the LHS of the above inequality to one, which gives $E^*\simeq 10$ MeV for both species. Such a rough estimate seems to work better for protons than for electrons as we have not taken into account the inhomogeneity of source distributions. We note also that $E^*$ would be larger for a smaller source rate $\mathcal{R}_s$. 

Concerning the stochastic fluctuations, it is also straightforward to see that above 100 MeV the uncertainty ranges are quite narrow since the energy loss time and the diffusive escape time are sufficiently large such that the distribution of CRs inside the Galactic disk become more or less uniform. 
We could see that the uncertainty ranges increase at lower energy until a characteristic energy $E^*$ below which the ratio between the upper and lower limit of the intensities becomes constant (the slopes of the intensities are approximately the same in all the realizations considered at energy $E<E^*$ as mentioned above). Since these fluctuations of the intensities are largest in the MeV energy range, the corresponding values of the ionization rate might also vary significantly from one realization to another and, thus, it might be hard to provide a single representative value for the ionization rate given our lack of knowledge on the exact locations and ages of CR sources. We might instead predict the \textit{range} of values for the ionization rate.  

More importantly, we also see in Fig.~\ref{fg:stochastic} that the intensities of CRs at the SPA position (right panel) is higher than the ones at the LOC position (left panel), especially at low energies. This is because the SPA point is located in regions with more CR sources in its vicinity. Since low-energy CRs are believed to be the main ionizing agents in MCs, it is also clear from this result that MCs positioned within spiral arms or, more generally, regions with high density of sources might have much higher ionization rates. We suggest that the difference in source density at different positions in the Milky Way together with the stochastic fluctuations might help to explain not only the surprisingly high ionization rate in some MCs but also the variations of this rate by more than one order of magnitude for clouds with similar column density.

\section{Cosmic-Ray Transport in Molecular Clouds}
\label{sec:cloud}
Before proceeding to the results on the stochastic fluctuations of the ionization rates in MCs, a more detailed discussion on CR transport into MCs is in order. The transport of CRs into MCs shall be analyzed using 1D models, where CRs are assumed to propagate only along magnetic field lines. Such a 1D description is sufficient provided that (i) the propagation of particles across magnetic field lines is unimportant, and (ii) the spatial scales relevant to the problem are smaller than, or at most comparable to the magnetic field coherence length in the ISM (typically assumed to be $l_c\sim 50$ - $100$ pc). Since these conditions are believed to be often satisfied, the 1D setup has been commonly adopted in the past literature to describe the penetration of CRs into MCs \citep{skilling1976,cesarsky1978,morfill1982,everett2011,morlino2015,schlickeiser2016,ivlev2018,phan2018}. In the following, we shall consider the two extreme cases of CR transport into MCs which are the ballistic model \cite{padovani2009} and the diffusive model (or more precisely the self-regulating diffusive transport model \cite{morlino2015,phan2018}).
Roughly speaking, if we consider a cloud of size $L_{cl}$, these models correspond respectively to the limits $l_c\simeq L_{cl}$ (ballistic) and $l_c\gg L_{cl}$ (diffusive) which could be translated into different boundary conditions for the problem. We shall now briefly review some of the key equations for the two models which will be adopted later for the analysis of the ionization rate. Interested readers might find a more thorough discussion on these models in the recent review by \cite{gabici2022}.

\begin{figure}[htpb]
\includegraphics[width=3.4in, height=1.6in]{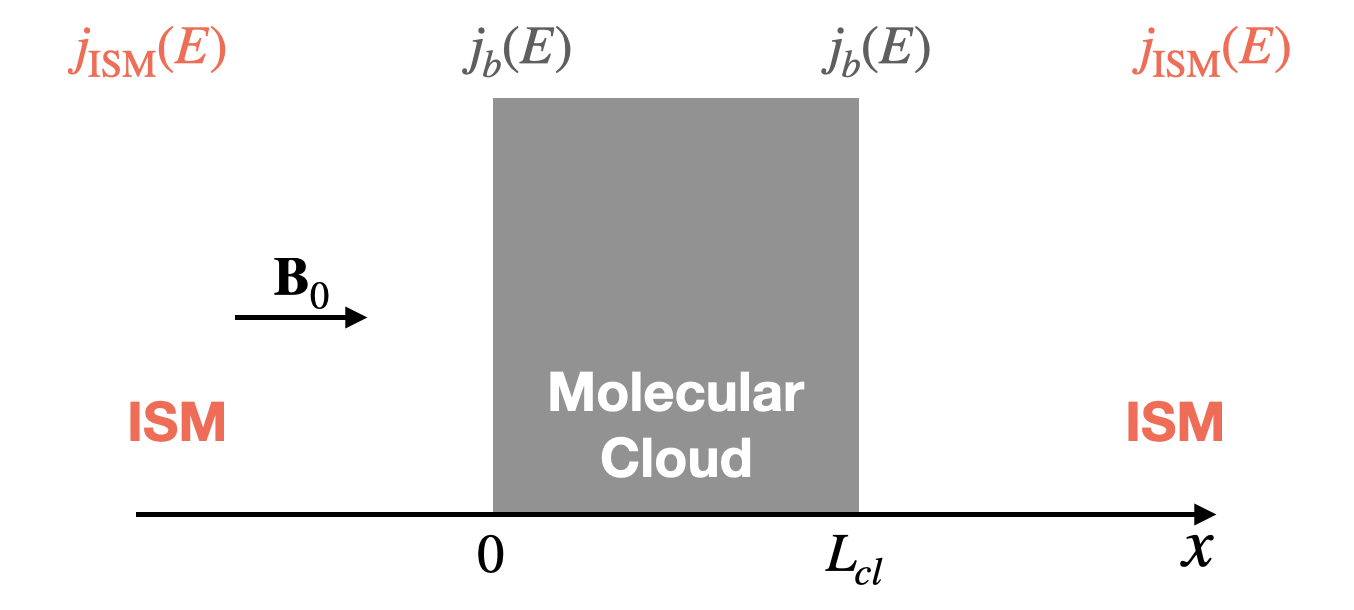}
\caption{Schematic view for the one-dimensional transport models describe in Section \ref{sec:cloud}. The shaded area represents the molecular cloud made up by molecular hydrogen. The ordered magnetic field threading the cloud is denoted as ${\bf B}_0$. The $x$-direction is chosen to be along ${\bf B}_0$. We denote the CR intensity at the cloud border and in the ISM respectively as $j(x=0,E)=j(x=L_{cl},E)=j_b(E)$ and $j(x\rightarrow \pm\infty,E)=j_{\rm ISM}(E)$ (see text for more details).  
}
\label{fg:cloud-sketch}
\end{figure}

In both models, the propagation of CRs inside the cloud is assumed to be ballistic as magnetic turbulence is likely to be dissipated by the ion-neutral damping \cite{zweibel1982}. Let's now choose the coordinate system as indicated in Fig.~\ref{fg:cloud-sketch} and adopt the \textit{continuously slowing down approximation} as done in Ref \cite{padovani2009}. This should allow us to relate the energy $E$ of a particle at a position $x$ inside the cloud to the \textit{initial energy} that the particle had when entering the cloud. We shall assume also the symmetry condition meaning the intensity of CRs on the left and right side of the cloud is identical. The intensity at a position $x$ inside the cloud can then be written as    
\begin{eqnarray}
j(x,E)=\frac{1}{2}\left[j_b(E_{01})\frac{\dot{E}_{cl}(E_{01})v}{\dot{E}_{cl}(E)v_{01}}+j_b(E_{02})\frac{\dot{E}_{cl}(E_{02})v}{\dot{E}_{cl}(E)v_{02}}\right]\n\\
\label{eq:jxE}
\end{eqnarray}
where $j_b(E)\equiv j(x=0,E)=j(x=L_{cl},E)$ is the intensity of CRs at the cloud border, $\dot{E}_{cl}$ is the energy loss rate due to interaction of CRs with molecular hydrogen inside clouds adopted from Ref. \cite{padovani2009}, $E_{01}=E_0(x,E)$ and $E_{02}=E_0(L_{cl}-x,E)$ are the initial energies for particles entering from the left and right edge of the cloud, and $v_{01}$ and $v_{02}$ are the corresponding speed for particles with energies $E_{01}$ and $E_{02}$. Here, we have introduced also the function $E_0(x,E)$ for the initial energy which is the solution of the following integral equation \cite{padovani2009,phan2018}
\begin{eqnarray}
x=\frac{1}{2}\int^E_{E_0}\frac{v}{\dot{E}_{cl}}\df E.
\end{eqnarray}
Note also that, in order to estimate the ionization rate, we normally adopt the intensity of CRs averaged over the whole MC meaning
\begin{eqnarray}
j_a(E)=\frac{1}{L_{cl}}\int^{L_{cl}}_0j(x,E)\df x.
\label{eq:ja}
\end{eqnarray}

At this point, it is worth introducing also the intensity of CRs in the surrounding ISM which is the boundary condition of the problem and could be formally defined as $j_{\rm ISM}(E)\equiv j(x=\pm\infty,E)$. The difference between the two models arises from the relation between $j_{\rm ISM}(E)$ and $j_b(E)$.

\subsection{Diffusive Model}

In this model, the presence of the cloud introduces a gradient in the CR intensity from outside to inside the cloud, due to severe energy losses operating in the cloud interior. Such an inhomogeneous distribution of CRs might excite magnetic turbulence in the cloud vicinity via the so-called \textit{streaming instability} \cite{,wentzel1974,skilling1976,morlino2015,schlickeiser2016,ivlev2018,marcowith2021,jacobs2022,gabici2022}. 
Magnetic turbulence excited in the surroundings of the cloud could then prevent particles from entering the cloud's interior and, thus, the CR intensities at the edge  and inside of the clouds might be suppressed in comparison to the ISM ones. The framework for the transport of CRs in this scenario has been developed in Ref. \cite{morlino2015} and \cite{phan2018} where the authors have shown that $j_b(E)$ and $j_{\rm ISM}(E)$ are related as follows 
\begin{eqnarray}
j_b(E)=\frac{j_{\rm ISM}(E)+\dfrac{v^2\dot{E}_{cl}(E^{\rm max}_0)}{4v_Av_{0}^{\rm max}\dot{E}_{cl}(E)}j_b(E^{\rm max}_{0})}{1+\frac{v}{v_A}},
\label{eq:jb-diff}
\end{eqnarray}
where $v_A\simeq 200$ km/s is the Alfv\'en speed at the cloud border assuming the ion density to be roughly $10^{-2}$ cm$^{-3}$ and the magnetic field strength in the vicinity of the cloud to be $B_0\simeq 10$ $\mu$G, $E^{\rm max}_0=E_0(L_{cl},E)$ and $v_0^{\rm max}$ is the corresponding speed of particles with energy $E^{\rm max}_0$. Strictly speaking, Eq. \ref{eq:jb-diff} does not give the exact form of $j_b(E)$ as it requires $j_b(E_0^{\text{max}})$ which, in principle, is unknown. However, we know that very high-energy particles, e.g. $E\gg 10$ GeV, could easily penetrate the cloud without suffering significant energy loss. This means that we could start from an arbitrarily large energy assuming $j_b(E)\simeq j_{\rm ISM}(E)$ for $E\gg 10$ GeV and adopt Eq. \ref{eq:jb-diff} to find $j_b(E)$ at smaller energies. We refer interested readers to Ref. \cite{phan2018} for more discussions on the numerical method to relate $j_b(E)$ and $j_{\rm ISM}(E)$ in the diffusive model.

\subsection{Ballistic Model}

The boundary condition in this case is imposed such that
\begin{eqnarray}
j_b(E)=j_{\rm ISM}(E)
\end{eqnarray}
This means that the CR induced magnetic turbulence in the cloud vicinity should be much weaker in order for this boundary condition to apply. Such a scenario might be realized if the correlation length of the magnetic field is roughly the same as the size of the cloud $l_c\simeq L_{cl}$. This is because the short correlation length might result in the field line wandering in 3D that reduces the strength of the CR gradient in comparison to the gradient in the 1D setup.   

In a sense, the ballistic and diffusive model are two extreme limits for the transport of CRs into MCs \cite{gabici2022}. More importantly, the two models will provide identical results for clouds of very low column density. This is because for very diffuse clouds the energy loss of CRs in the penetration process becomes negligible even in the MeV energy range and, thus, the self-generated magnetic turbulence at the cloud border is no longer effective.     

At this point, we would like to open a parenthesis and provide a more technical discussion concerning an alternative diffusive model which has been developed in Ref. \cite{ivlev2018}. In this case, the authors also involve the self-generated turbulence of CRs in describing the transport into MCs. This model, however, assumes also $j_b(E)=j_{\rm ISM}(E)$ and takes into account the streaming instability only inside clouds. A more thorough comparison between this model and the ballistic model has been provided in Ref. \cite{silsbee2019} assuming different parametric models for the CR intensities at low energies. More importantly, it has been pointed out that the diffusive motion of particles inside clouds might also depend on the ion fraction within these clouds which might be quite uncertain and could vary on a case by case basis. For this reason, we shall limit ourselves to the ballistic and our version of the diffusive model for computing the ionization rate. More investigations on all of these models taking into account chemical properties of clouds and magnetic field geometry shall be carried on in future works.   

Once $j_b(E)$ is known for a given $j_{\rm ISM}(E)$ and a transport model, Eq. \ref{eq:jxE} and \ref{eq:ja} could be used to solve for the average intensity for any MCs of interest. As mentioned above, most of the previous works have only adopted the intensity observed locally (e.g. by Voyager and AMS) or some extrapolated versions of the local intensity (see e.g. the \textit{high} model in Ref. \cite{ivlev2015}) for $j_{\rm ISM}$ to estimate the ionization rate. We shall show in the next section that using $j_{\rm ISM}(E)$ at different positions in the Milky Way and also taking into account fluctuations among various realizations of the CR sources' distribution in space and time might lead to much higher ionization rates that could fit better the observed data in some cases.

\section{Stochastic Fluctuations of the Ionization Rate}
\label{sec:ionization}
Having discussed the transport model of CRs into MCs, we are in a position to examine the stochasticity of the ionization rate. This should be done by first evaluating the intensities of CRs inside an arbitrary MC for either the ballistic or diffusive model (as introduced in the previous section) with the intensity in the ISM given as $j_{{\rm ISM},s}(E)=j^{(n)}_{s}(E)$ where $s$ represents the species (proton or electron) and $j_{s}^{(n)}(E)$ is the intensity of species $s$ in the $n$th realization. It is then quite straightforward to estimate the respective ionization rate induced by both CR protons and electrons for each realization to derive the most probable range for the values of $\zeta({\text{H}_2})$. 

The ionization rate can be evaluated as follows \citep{padovani2009,chabot2016,recchia2019,phan2018,gabici2022}
\begin{eqnarray}
&&\zeta_p(\text{H}_2)=\int^{E_{\text{max}}}_I 4\pi j_{a,p}(E)\left[1+\phi_p(E)\vphantom{^{\frac{•}{•}}}\right]\sigma^p_{\text{ion}}(E)\df E\n\\ 
&&\qquad\qquad\qquad+\int^{E_{\text{max}}}_0 4\pi j_{a,p}(E) \sigma_{\text{ec}}(E)\vphantom{^{\frac{a^b}{•}}}\df E\label{eq:ion_p}\\
\n\\
&&\zeta_e(\text{H}_2)=\int^{E_{\text{max}}}_I 4\pi j_{a,e}(E) \left[1+\phi_e(E)\vphantom{^{\frac{•}{•}}}\right]\sigma^e_{\text{ion}}(E)\vphantom{^{\frac{a^b}{•}}}\df E\n\\
\label{eq:ion_e}
\end{eqnarray}
where $j_{a,p}(E)$ and $j_{a,e}(E)$ are respectively the intensities of CR protons and electrons spatially averaged over the whole MC, $\sigma^p_{\text{ion}}$, $\sigma_{\text{ec}}$, and $\sigma^e_{\text{ion}}$ are the proton ionization cross section, the electron capture cross section, and the electron ionization cross section, respectively. The functions $\phi_p(E)$ and $\phi_e(E)$ represents the average secondary ionization per primary ionization due to CR protons and electrons respectively as in \cite{krause2015,gabici2022} (see also \cite{ivlev2021} for more discussions). The ionization potential of H$_2$ is taken to be $I\simeq15.426$ eV. It should be noted that the ionization rate for all nuclei has been obtained by multiplying the one for protons with the nuclear enhancement factor $\eta\simeq 1.5$ and, thus, the total ionization rate is $\zeta(\Hm)=\eta\zeta_p(\Hm)+\zeta_e(\Hm)$. Even though the stochastic fluctuations have been investigated for CRs of energy in the range from $1$ MeV to $10$ GeV, we will derive the ionization rate from particles with energy above $1$ keV as sub-MeV particles might provide a non-negligible contribution to the ionization rate, especially for CR electrons (see e.g. \cite{padovani2009,ivlev2015}). To this end, we have extrapolated the upper and lower limit intensities down to 1 keV employing the fact that $j(E)\sim v/\dot{E}$ for $E\lesssim E^*$. 

\begin{figure*}[htpb]
\includegraphics[width=3.5in, height=2.8in]{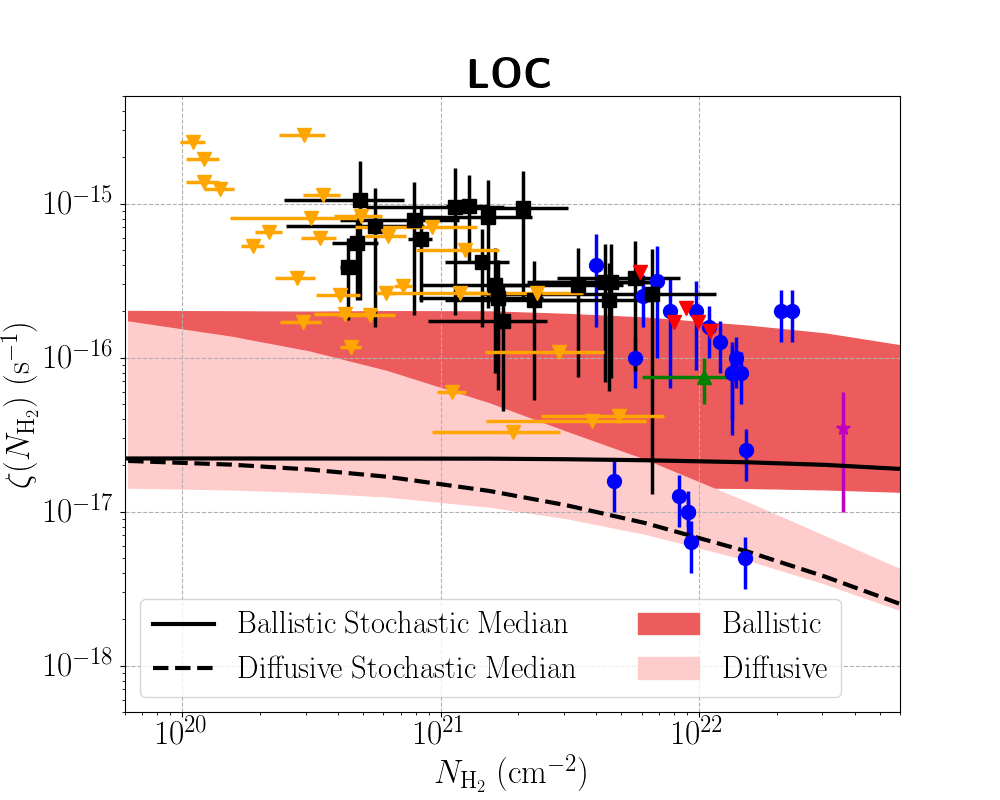}
\includegraphics[width=3.5in, height=2.8in]{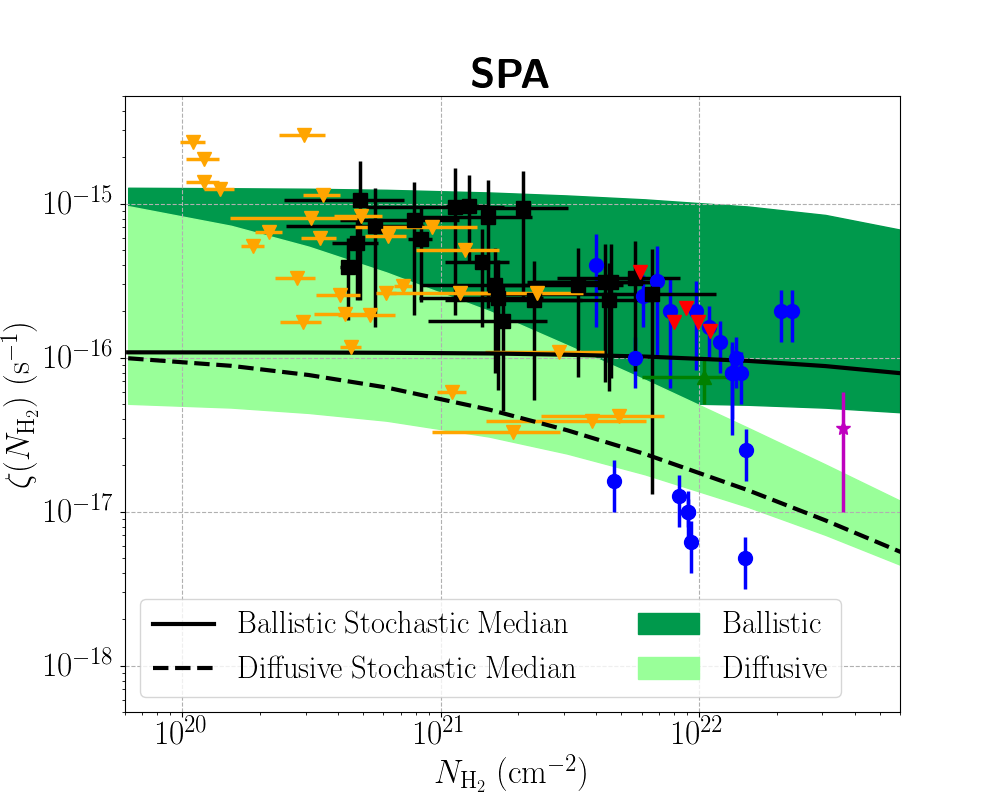}
\caption{Stochastic fluctuations of the ionization rate for the local ISM (left) and for the chosen point in a spiral arm (right). The dashed and solid black lines correspond to the median intensities predicted from the diffusive and ballistic model. Data for the ionization rate are from \cite{caselli1998} (filled blue circles), \cite{williams1998} (green triangle), \cite{bialy2022} (red triangles are upper limits), \cite{maret2007} (asterisk), and \cite{indriolo2012} (black squares are data points while inverted yellow triangles are upper limits).}
\label{fg:ion_SNR}
\end{figure*}

The statistical analysis of the ionization rate is also performed as in the case of the CR intensities (see Section \ref{sec:stochasticity}). We first identify the distribution function $p(\zeta)$ for the ionization rate at a column density and then characterize the most probable range of values for the ionization rate with the quantiles $\mathcal{I}_{95\%}=\left[ \zeta_{2.5\%},\zeta_{97.5\%}\right]$. The results can be fitted with the following expression
\begin{eqnarray}
\zeta=10^{a_0+a_1 Y+ a_2 Y^2 +a_3 Y^3} \, {\rm s}^{-1} \label{eq:eq-fit-ion}
\end{eqnarray}
where $Y=\log_{10}\left[N_{\rm H_2}/(6\times 10^{19}\,{\rm cm}^{-2})\right]$. We present the fit parameters in Table \ref{tab:fit-ion} for the potential range of the ionization rate together ($\zeta_{2.5\%}$ and $\zeta_{97.5\%}$) with the median values ($\zeta_{50\%}$). Note that we refer to $\zeta_{2.5\%}$, $\zeta_{50\%}$ and $\zeta_{97.5\%}$ respectively as the MIN, MED, and MAX limits in this table. Equation \ref{eq:eq-fit-ion} with these parameters provide handy formulae which could be useful for subsequent analyses on star formation or astrochemistry research.

\begin{table*}[!t]
\centering
\caption{Fit parameters for the stochastic fluctuations of the ionization rate in clouds.}

	\label{tab:fit-ion}
	\begin{tabular}{|c|c|c|c|c|c|c|} 
		\hline
		\hline
        Position & Model & Limit & $a_0$ & $a_1$ & $a_2$ & $a_3$ \\
        \hline
        \hline
        \multirow{6}{*}{LOC} & \multirow{3}{*}{Ballistic} & MIN &  $-1.68\times 10^{1}$ & $-8.85\times 10^{-3}$ & $1.22\times 10^{-2}$ & $-4.33\times 10^{-3}$ \\
        & & MED &  $-1.67\times 10^{1}$ & $-8.72\times 10^{-3}$ & $1.40\times 10^{-2}$ & $-6.23\times 10^{-3}$ \\
        & & MAX &  $-1.57\times 10^{1}$ & $-1.55\times 10^{-2}$ & $3.22\times 10^{-2}$ & $-1.74\times 10^{-2}$ \\
        \cline{2-7}
        & \multirow{3}{*}{Diffusive} & MIN &  $-1.68\times 10^{1}$ & $-4.50\times 10^{-3}$ & $-3.59\times 10^{-2}$ & $-1.78\times 10^{-2}$ \\
        & & MED &  $-1.67\times 10^{1}$ & $-2.52\times 10^{-2}$ & $-6.89\times 10^{-2}$ & $-9.30\times 10^{-3}$ \\
        & & MAX &  $-1.58\times 10^{1}$ & $-1.74\times 10^{-1}$ & $-1.62\times 10^{-1}$ & $1.21\times 10^{-2}$ \\
        \hline
        \hline
        \multirow{6}{*}{SPA} & \multirow{3}{*}{Ballistic} & MIN &  $-1.63\times 10^{1}$ & $-1.46\times 10^{-2}$ & $2.09\times 10^{-2}$ & $-7.96\times 10^{-3}$ \\
        & & MED &  $-1.60\times 10^{1}$ & $-1.32\times 10^{-2}$ & $2.00\times 10^{-2}$ & $-1.02\times 10^{-2}$ \\
        & & MAX &  $-1.49\times 10^{1}$ & $-2.47\times 10^{-2}$ & $2.33\times 10^{-2}$ & $-1.48\times 10^{-2}$ \\
        \cline{2-7}
        & \multirow{3}{*}{Diffusive} & MIN &  $-1.63\times 10^{1}$ & $-2.01\times 10^{-2}$ & $-8.07\times 10^{-2}$ & $-1.09\times 10^{-2}$ \\
        & & MED &  $-1.60\times 10^{1}$ & $-7.15\times 10^{-2}$ & $-1.21\times 10^{-1}$ & $7.73\times 10^{-4}$ \\
        & & MAX &  $-1.50\times 10^{1}$ & $-2.36\times 10^{-1}$ & $-2.22\times 10^{-1}$ & $2.94\times 10^{-2}$ \\
		\hline
		\hline
	\end{tabular}
\end{table*}

The stochastic fluctuations of the ionization rate (from both CR nuclei and electrons) for both the ballistic model and diffusive model, depicted respectively as the dark and light bands, are shown in Fig.~\ref{fg:ion_SNR} together with the measurements and upper limits of the ionization rate taken from \cite{caselli1998,williams1998,maret2007,indriolo2012,bialy2022}. Results are presented for both the local ISM and the chosen point in a spiral arm (LOC and SPA points as defined in Section \ref{sec:source-dist}). 

It is clear that the stochastic fluctuations could make the ionization rate vary by roughly more than one order of magnitude at both positions reaching the maximum of about $~2\times 10^{-16}$ s$^{-1}$ and $10^{-15}$ s$^{-1}$ respectively for the LOC and SPA points. We can also see that the ballistic and the diffusive model deviate quite significantly for high-column-density clouds. This is because the exclusion of CRs for the diffusive model seems to be so strong that the ionization rate decreases relatively quickly for increasing column density. Interestingly, such a discrepancy of the ionization rate between the two models matches well with the large variations in the observed values which might suggest that both models might be valid depending on the exact geometry of the magnetic field and the level of turbulence around these clouds. In fact, the CR intensities at the SPA point allow us to predict the most probable range of values for the ionization rate that fit quite well with data. 

We note however that the large discrepancy between the ionization rate expected with the local CR spectra and the observed values (also known as the ionization puzzle) is still not yet resolved. For diffuse clouds ($N_{\rm H_2}\lesssim 5\times 10^{21}$ cm$^{-2}$), even though all the data points and upper limits from are compatible (i.e. roughly within) with stochastic bands of ionization rates from both the ballistic and diffusive models, the mean value of ionization rate measurements (black data points from \cite{indriolo2012}) is actually roughly $3\times 10^{-16}$ s$^{-1}$ which is about 3 times higher than the median value expected in the most optimistic case namely the SPA ballistic model. Several clouds in the data set of \cite{indriolo2012} are also known to be quite nearby (as indicated by upper limits of their distances) and yet they also have relatively high ionization rates. Concerning dense clouds ($N_{\rm H_2}\gtrsim 5\times 10^{21}$ cm$^{-2}$), it is not straightforward to extract a mean value of the ionization rate from data as they vary rather strongly and might be sensitive to both column densities and the transport of CRs around these clouds. Given all these uncertainties, one might expect to make further progress by actually studying individual clouds where information about the geometry of the magnetic field and the level of turbulence are known with some confidence.

This result is based on the assumption that CR sources are more numerous within spiral arms than in the inter-arm region. 
Such an assumption, though plausible, currently lacks observational tests. 
For example, the spatial distribution of known SNRs does not seem to exhibit a spiral pattern \cite{ranasinghe2022}, but this might simply be due to the incompleteness of the sample.
In the same way, ionisation rates have been measured for molecular and atomic clouds located at various distances from the Sun \cite{indriolo2015}, but data are too sparse and uncertain to allow for claims on spatial correlations between enhanced ionisation rates and spiral arms.
Finally, and most importantly, large ionisation rates (up to the largest values of $\approx 10^{-15}$~s$^{-1}$) have been measured in a sample of nearby MCs, mostly located in the inter-arm region \cite{indriolo2012}.
These observations might be explained invoking either the presence of a local excess of CR sources, or a correlation between the position of MCs and CR sources (the latter is indeed expected for SNRs, see e.g. \cite{montmerle1979}.
Such scenarios will be investigated in a forthcoming publication.

This seems to indicate that these clouds are more likely to be in regions with higher source density than expected locally or there might be correlations between positions of MCs and sources.  

Another interesting remark is that, throughout this work, we have adopted an isotropic diffusion model. There exist also models where CRs have different diffusion coefficients parallel and perpendicular to Galactic magnetic field lines. These are referred to as anisotropic diffusion models characterized by $D^{\rm ani}_{\perp}(E)$ and $D^{\rm ani}_{\perp}(E)$. In anisotropic diffusion models, the escape of CRs from the disk around the location of the Solar System is normally determined by $D^{\rm ani}_{\perp}(E)$ \cite{evoli2012,cerri2017}. Thus, we should have $D^{\rm ani}_{\perp}(E)=D(E)$ (where $D(E)$ is the isotropic diffusion coefficient that we adopt above) in order to have the same spectral index for the local CR spectra in both anisotropic and isotropic diffusion models. This means that the diffusion loss lengths perpendicular and parallel to the Galactic magnetic field in anisotropic diffusion models should be $l_\perp^{\rm ani}(E)\simeq l_d(E)\simeq \sqrt{4D(E)\tau_l(E)}$ and $l_\parallel^{\rm ani}(E)>l_d(E)$. This will lead to the characteristic energy $E^*_{\rm ani}$ at which stochasticity is maximized to be smaller in anisotropic diffusion models $E^*_{\rm ani}<E^*$ (see Eq. \ref{eq:Estar}). As a result, the median spectra in anisotropic diffusion models might have more CRs of energy $E<E^*$ than the one in isotropic diffusion models to induce more ionizations. However, since CRs in anisotropic diffusion models are more dispersed due to $l_\parallel^{\rm ani}(E)>l_d(E)$, we expect the stochastic band to be more narrow. Given these qualitative expectations on the CR spectra, the corresponding ionization rates in anisotropic diffusion models might have a larger median value but smaller uncertainty bands. In this respect, stochastic fluctuations in anisotropic diffusion models with the Galactic magnetic field taken into account will be interesting also for future investigations.

\begin{figure*}[htpb]
\includegraphics[width=3.5in, height=2.8in]{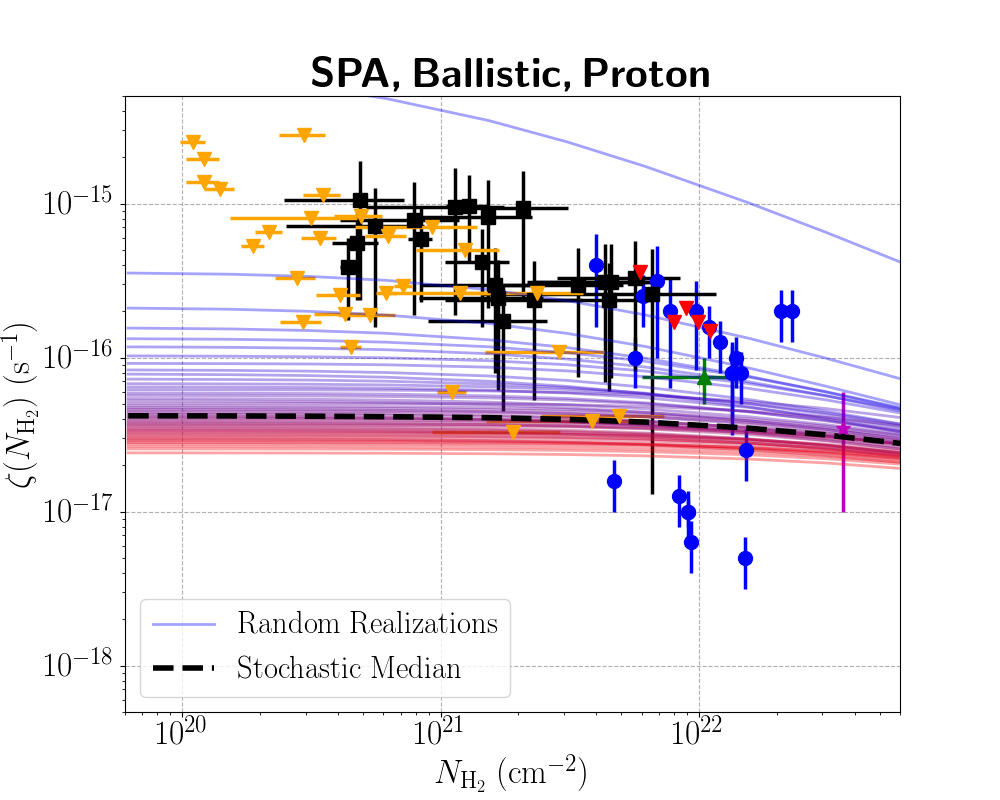}
\includegraphics[width=3.5in, height=2.8in]{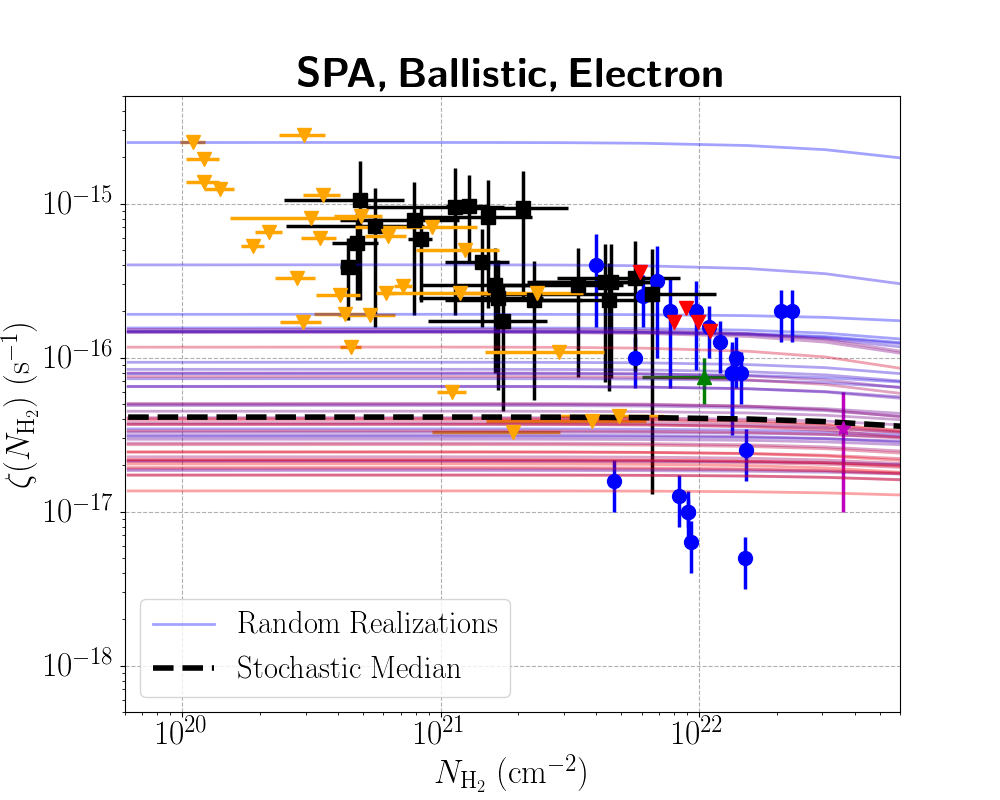}\\
\includegraphics[width=3.5in,height=2.8in]{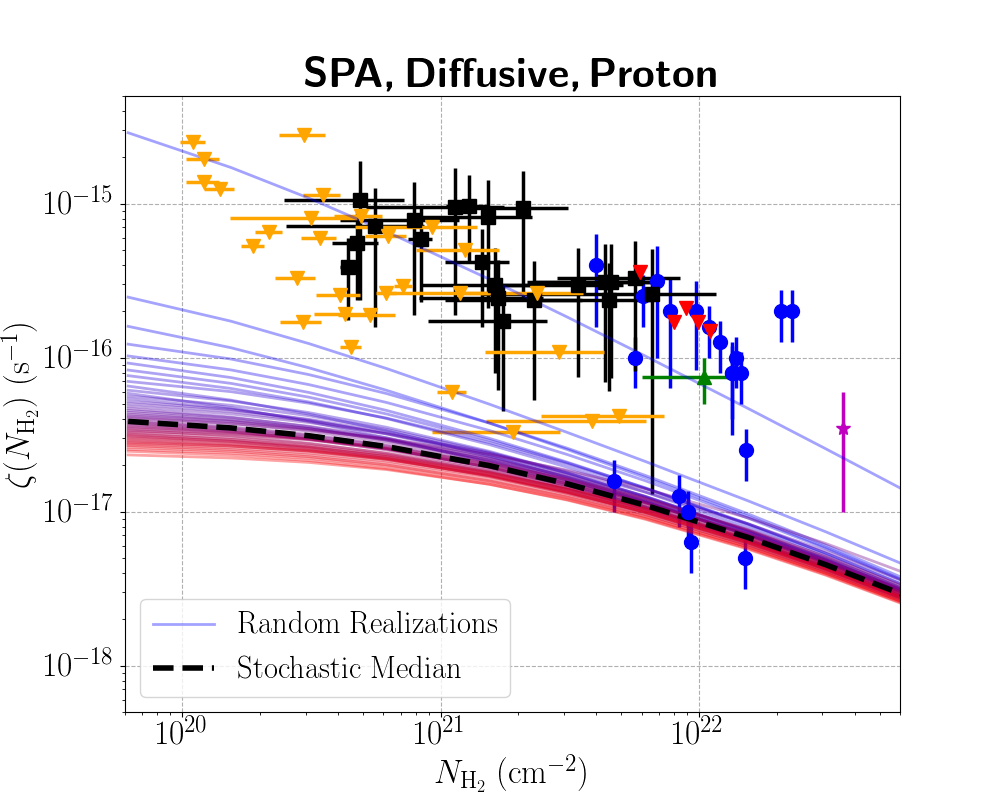}
\includegraphics[width=3.5in, height=2.8in]{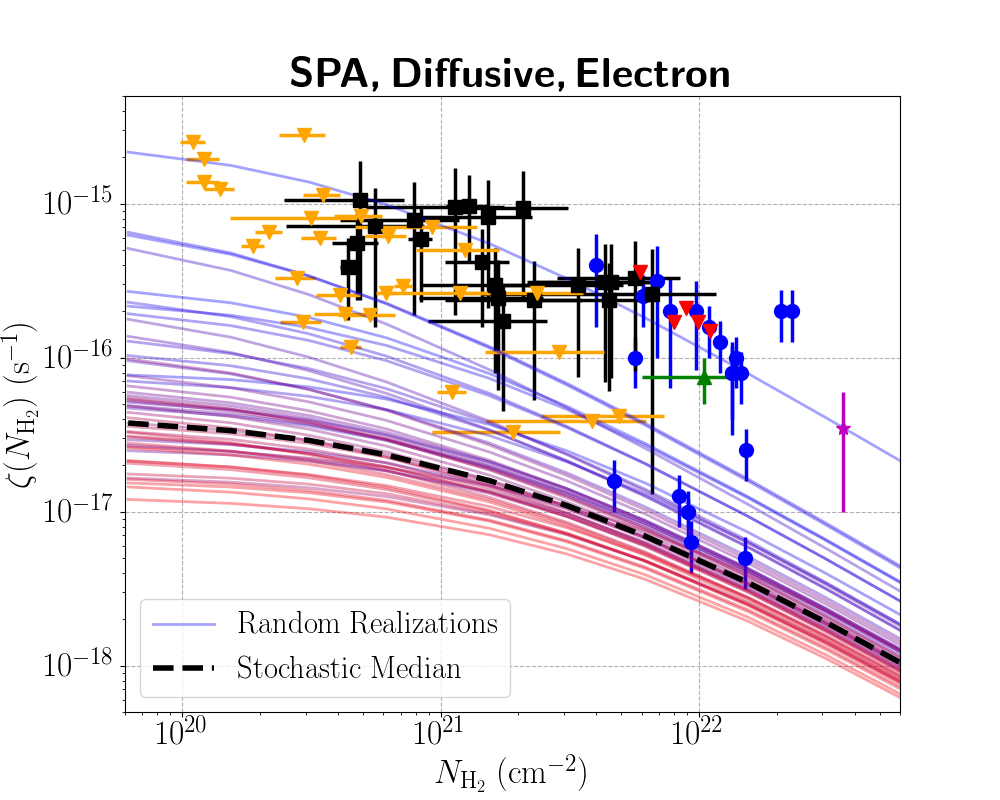}
\caption{Ionization rate in 50 simulated realizations of sources around the SPA point, represented by solid lines, from CR protons (left column) and CR electrons (right column). Note that solid lines with the same shade of color mean that the result have been derived with the same realization of SNRs. The dashed black lines are the stochastic median of the ionization rate values. Data for the ionization rate are from \cite{caselli1998} (filled blue circles), \cite{williams1998} (green triangle), \cite{bialy2022} (red triangles are upper limits), \cite{maret2007} (asterisk), and \cite{indriolo2012} (black squares are data points while inverted yellow triangles are upper limits).}
\label{fg:ion_pe}
\end{figure*}

We present also in Fig. \ref{fg:ion_pe} the ionization rate induced by CR protons (left) and electrons (right) from 50 random realizations of SNRs at the SPA point for both the ballistic and diffusive models (upper and lower panels respectively). For the ballistic model, the ionization rate is roughly independent of the column density and the stochastic medians are roughly equal for both species. For the diffusive model, the ionization rate decreases quickly with column density and the stochastic medians are also roughly the same at low column density. In case of dense clouds ($N_{\rm H_2}\gtrsim 10^{22}$ cm$^{-2}$), the stochastic median of nuclei seems to be slightly larger than that for electrons. Note however that the stochastic medians only provide rough estimate for the contribution of each species in an average sense and the relative contribution of each species to the total ionization rate might change significantly from one realization of SNRs to another. It is interesting to remark also that the stochastic fluctuations of the ionization rate induced by electrons seem to be stronger than that for protons in both the ballistic and diffusive models. This can be seen in Fig.~\ref{fg:ion_pe}, where the ionization rate produced by electrons and protons for a given realization of SNRs is represented with the same shade of color in the left (protons) and right (electrons). The spread in ionization rate in the case of electrons appears markedly larger than that induce by protons

\section{Summary and conclusions}
\label{sec:conclusion}

It is clear that CRs play an essential role in determining the physical and chemical evolution of star-forming regions as they are capable of penetrating and ionizing the interior of MCs. This means that the impact of CRs on MCs could be quantified by the CR induced ionization rate. Interestingly, theoretical estimates of the ionization rate assuming the CR spectra observed in the local interstellar medium normally result in the ionization rate being one to two orders of magnitude below the values inferred from observations (sometimes referred to as \textit{the ionization puzzle}). Such a discrepancy might be resolved by arguing that the local CR spectra are not representative for the entire Galaxy. Here, we model the distribution of low-energy CR spectra expected from a statistical model of discrete sources at different positions in the Milky Way. The corresponding distribution for the ionization rate is then derived and confronted with data to demonstrate how the variations of CR density at low energies might help to explain the scattered values and the surprisingly high ionization rates observed in many molecular clouds.

To this end, we follow the framework presented in Ref. \cite{phan2021} to model the CR spectra in the energy range from 1 MeV to 10 GeV assuming SNRs are the main sources of these particles. We first generate many different realizations of source ages and distances and then estimate the CR spectra contributed from all the sources within each realization. This gives us an ensemble of the CR spectra from which one could predict the theoretical uncertainty of the spectra. Such a theoretical uncertainty (also referred to as stochastic fluctuations) reflects the potential variations of CR density on small scales (scales smaller or roughly equal to the mean distance between sources or the diffusion loss length of MeV CRs). On larger scales, e.g. from interarm to spiral-arm regions, the mean source density might also vary and, thus, the CR spectra are expected to change significantly, especially for MeV CRs. In order to gain more insight into the fluctuations both on large and small scale, we study the CR spectra for two representative positions in the Milky Way namely the LOC and SPA points as defined in Section \ref{sec:source-dist}. We found that stochastic fluctuations are indeed relevant in the MeV energy range and the SPA point, in general, have larger CR density than the LOC point due to the higher source density within spiral arms.  

In order to study the ionization rate, we have to specify also models to describe the penetration of CRs into MCs. There exists many models in the literature which might be applicable depending on the properties of MCs and the Galactic magnetic field. Here, we adopt the two models following Ref. \cite{padovani2009} and \cite{phan2018}  referred to as the ballistic and diffusive models. Using the CR spectra in all realizations together with the two transport models into clouds, we have estimated the ensemble of ionization rates for the two representative points. The main results could be summarized as follows

i) The stochastic fluctuations of the CR spectra (or equivalently variations on small scales) mean that we could only predict a range of values for the ionization rate. This might help to explain the large scatter in the measured values of the ionization rate. 

ii) Even if we take into account the stochasticity of the ionization rate, the source density in the local ISM (LOC point) is still not sufficient to explain the observed data. Larger values of the ionization rate might be achievable if clouds are located in regions of higher source density, e.g. within spiral arms, as in the case of the SPA point considered in this paper. However, quite large values of the ionisation rate have been measured also in MCs located in the inter-arm region. Explaining this measurements requires either the presence of a local excess of CR sources, or a correlation between the positions of MCs and CR sources. These scenarios will be described in a separate publication.

iii) We also find that the difference in the predicted range of ionization rates could be more than one order of magnitude for dense clouds ($N_{\textrm{H}_2}\gtrsim 10^{22}$ cm$^{-2}$) depending on the model of CR transport into clouds. This is because the suppression of the CR intensities as particles propagate from outside to inside dense clouds might change significant for different correlation lengths of the magnetic field threading these clouds. If we consider only measurements and not upper limits, there seems to be a large scatter for the observed data in accordance with our expectation, see e.g. the blue data points from \cite{caselli1998} in Fig. \ref{fg:ion_SNR} with the ionization rate values differ by more than one order of magnitude for clouds with rather similar column density. 

As a final remark, we note that all the ranges of values for both the representative points in the Milky Way (LOC and SPA) using different models of CR transport into clouds are reported Table \ref{tab:fit-ion}. This might serve as a handy look-up table for the values of the ionization rate for both the Astrochemistry and Star Formation Communities.  

\begin{acknowledgments}
This project was funded by the Deutsche Forschungsgemeinschaft (DFG, German Research Foundation) -- project number 490751943 and by Agence Nationale de la Recherche (project CRitiLISM, ANR-21-CE31-0028). 
The work of SR  is partially supported by the {\sc Departments of Excellence} grant awarded by the Italian Ministry of Education,
University and Research ({\sc Miur}), the 
Research grant {\sl The Dark Universe: A Synergic Multimessenger Approach}, No.
2017X7X85K funded by the {\sc Miur} and by the 
Research grant {\sc TAsP} (Theoretical Astroparticle Physics) funded by Istituto
Nazionale di Fisica Nucleare. 
\end{acknowledgments}

\bibliographystyle{apsrev4-2}
\bibliography{mybib}

\end{document}